\documentclass{IEEEtran}
\pdfoutput=1
\usepackage{amsmath,amssymb,amsfonts}
\usepackage{algorithm}
\usepackage{algorithmicx}
\usepackage{algpseudocode}
\usepackage{graphicx}
\usepackage{textcomp}
\usepackage{subcaption}
\usepackage{multicol}
\usepackage{lipsum}
\usepackage{enumerate}
\usepackage{bm}
\usepackage[numbers]{natbib}
\newtheorem{lemma}{Lemma}
\newtheorem{theorem}{Theorem}
\newtheorem{corollary}{Corollary}

\newcommand{\RNum}[1]{\uppercase\expandafter{\romannumeral #1\relax}}

\def\BibTeX{{\rm B\kern-.05em{\sc i\kern-.025em b}\kern-.08em
    T\kern-.1667em\lower.7ex\hbox{E}\kern-.125emX}}
\begin{document}

\title{Reaching Efficient Byzantine Agreements in Bipartite Networks
\thanks{This work has been submitted to the IEEE for possible publication. Copyright may be transferred without notice, after which this version may no longer be accessible.}
}

\author{Shaolin Yu*, Jihong Zhu, Jiali Yang, Wei Lu\\Tsinghua University, Beijing, China \\ysl8088@163.com}

\maketitle

\begin{abstract}
For reaching efficient deterministic synchronous Byzantine agreement upon partially connected networks, the traditional broadcast primitive is extended and integrated with a general framework.
With this, the Byzantine agreement is extended to fully connected bipartite networks and some bipartite bounded-degree networks.
The complexity of the Byzantine agreement is lowered and optimized with the so-called Byzantine-levers under a general system structure.
Some bipartite simulation of the butterfly networks and some finer properties of bipartite bounded-degree networks are also provided for building efficient incomplete Byzantine agreement under the same system structure.
It shows that efficient real-world Byzantine agreement systems can be built with the provided algorithms with sufficiently high system assumption coverage.
Meanwhile, the proposed bipartite solutions can improve the dependability of the systems in some open, heterogeneous, and even antagonistic environments.
\end{abstract}


\section{Introduction}
\label{sec:Introduction}
Computation and communication are two basic elements of distributed computing and are fundamental in constructing distributed real-time systems.
In real-world systems, these two elements are often modularized into \emph{computing components} and \emph{communicating components}.
Both these components can be with failures that should be tolerated in reliable systems.
For tolerating malign faults, the Byzantine agreement (BA, also \emph{Byzantine Generals}) problem has drawn great attention in both theoretical and industrial realms \citep{RN2119,RN4551}.
Theoretically, solutions of the BA problem are proposed with authenticated messages \citep{RN2119,RN4137,Abraham2019}, exponential information gathering (EIG) \citep{RN4250,Garay2006Fully,Kowalski2013Synchronous,RN4148}, unauthenticated broadcast primitive \citep{SrikanthSimulating,Toueg1987Fast}, phase-king (and queen) \citep{Berman1989optimalconsensus}, and other basic strategies.
However, reliable authentication protocols are inefficient and impractical in most real-time distributed systems.
Meanwhile, most of the unauthenticated BA solutions require fully connected networks, which can hardly be supported in practical distributed systems.
Also, the computation, message, and time complexities required in classical BA algorithms are often prohibitively high.
As a result, the Byzantine fault-tolerant (BFT) mechanisms employed in real-world systems often take some local \emph{Byzantine filtering} schemes \citep{RN4551,as6003,as6802} which have to be very carefully designed, implemented, and verified in providing adequate assumption coverage \citep{Powell1997Coverage,Kopetz2004Hypothesis}.
For avoiding this, a fundamental problem is how to reach efficient BA in partially connected networks without authenticated messages nor local \emph{Byzantine filtering} schemes.

\subsection{Motivation}
In this paper, we investigate the BA (and almost everywhere BA \citep{Dwork1986}) problem upon bipartite networks and aim to provide efficient solutions.
Firstly, with the generalized model \cite{ASBBDNFTP} of the relay-based broadcast systems, we discuss how to efficiently extend the classical broadcast primitive \citep{SrikanthSimulating} to fully connected bipartite networks.
Then, we construct a modularized BA system by utilizing the broadcast systems as the corresponding broadcast primitives upon networks of general topologies.
Concretely, as the provided BA system structure is independent of network topologies, the proposed broadcast systems can be integrated with a general BA framework \citep{RN4346,Toueg1987Fast} for reaching efficient BA upon networks of arbitrary topologies.
Meanwhile, based upon our previous work \cite{ASBBDNFTP} which only handles the broadcast problem with non-bipartite bounded-degree networks, here we also discuss how to extend the solutions to bipartite bounded-degree networks.
For this, firstly, we explore the almost everywhere BFT solutions with bipartite simulation of the classical butterfly networks \citep{Dwork1986}.
Then, we also explore some basic properties in generally extending the almost everywhere BFT solutions upon non-bipartite networks to the corresponding ones upon bipartite networks.
Lastly, we investigate the system assumption coverage of the BFT systems upon bipartite networks and discuss how these systems can outperform the ones built upon non-bipartite networks.

\subsection{Main contribution}
Firstly, by reviewing the traditional broadcast primitive with the general system model, some necessary conditions and sufficient conditions for solving the broadcast problem are formally derived.
With this, efficient broadcast solution upon fully connected bipartite networks is provided.

Secondly, in providing broadcast solutions upon bipartite bounded-degree networks, the bipartite simulation achieves similar efficiency of the original butterfly-network solution.
Meanwhile, by extending some classical result of non-bipartite bounded-degree networks to bipartite bounded-degree networks, several classical almost everywhere BFT solutions upon non-bipartite networks can be easily applied to the corresponding bipartite networks.

Thirdly, the modularized BA solution gives a general way to build efficient BA systems upon arbitrarily connected networks.
Meanwhile, by exploiting the differences of the componential failure rates in bipartite networks, the efficiency of BA is improved in comparing with the classical deterministic BA upon fully connected networks where the componential failure rates can only be assumed with the most unreliable components of the systems.

\subsection{Some benefits}
This work may have several benefits.
Firstly, the BA problem upon bipartite networks (referred to as the BABi problem) can draw a closer relationship between theoretical BFT solutions and practical fault-tolerant systems.
Theoretically, as is shown in \cite{RN4347}, the network connectivity required in solving the BA problem without authentication cannot be less than $2f+1$ where $f$ is the maximal number of Byzantine nodes.
Among all available topologies, the bipartite graph is the simplest way to break the complete graph assumption while maintaining the required connectivity for reaching a complete BA.
Practically, as bipartite topologies are easy to be implemented with redundant bus-based or switch-based communication networks, they can be conveniently adopted in building real-world small-scale systems.

Secondly, for large-scale systems, by extending the almost everywhere BFT solutions upon non-bipartite networks to the ones upon bipartite networks, the large-scale systems can be better designed with layered architectures.
As the distributed components in the same side of the bipartite network need not to directly communicate with each other, these components can be heterogeneous, free of relative physical location restrictions, and unknown to each other.
This not only facilitates the deployment of the different layers that might be operated by lots of people and different organizations but saves directional communication resources in some emerging systems such as the multi-layer satellite networks.

Thirdly, the BABi algorithms designed for systems with programmable communicating components can be more efficient than the traditional BA algorithms designed only for computing components.
Also, as the failure rates of different kinds of components in the system can be significantly different, the BABi solutions can make a leverage on the two sides of the bipartite networks.

Lastly, by providing efficient BA solutions upon bipartite networks, many other related BFT problems can be handled more practically.
For example, the provided BABi solutions can help to build self-stabilizing BFT synchronization \cite{daliot2006self,RN4069,TDWALDEN} in systems with bounded clock drifts, bounded message delays, and bounded node-degrees.

\subsection{Paper layout}
The structure of the paper is sketched as below.
Firstly, the related work is presented in Section~\ref{sec:Related}, with the emphasis on the BFT systems provided upon partially connected networks.
The most general system settings and the basic problem definitions are presented in Section~\ref{sec:Model}, where the system, the nodes (correct and faulty ones), the synchronous communication network, the strong (but not adaptive) adversary, and the general problems are given.
Then, we discuss the broadcast systems upon the fully connected networks, the fully connected bipartite networks, and the bipartite bounded-degree networks in Section~\ref{sec:Broadcast_1}, Section~\ref{sec:Broadcast_2}, and Section~\ref{sec:Broadcast_3}, respectively.
We will see that, as the traditional restriction on the broadcast system is relatively tight, the complete solutions (in Section~\ref{sec:Broadcast_1} and Section~\ref{sec:Broadcast_2}) are mainly explored in a restricted Boolean algebraic manner.
In the context of the incomplete solutions (in Section~\ref{sec:Broadcast_3}), the relaxed restriction makes rooms for richer studies.
Then in Section~\ref{sec:Agreement_Ex}, the general BA solution is provided under the modularized agreement system structure, the application of which is briefly discussed in Section~\ref{sec:app}.
In Section~\ref{sec:Agreement_Ex} and Section~\ref{sec:app}, the efficiency of the provided BABi solution is also discussed, with which we show how the solutions upon the bipartite networks can break the classical BFT limitations upon real-world networks with sufficiently high system assumption coverage.
Finally, we conclude the paper in Section~\ref{sec:Conclusion}.

\section{Related work}
\label{sec:Related}
\subsection{The classical problem}
The Byzantine agreement (\emph{Byzantine Generals}) problem was introduced more than $40$ years ago in \cite{PSL1980} (\cite{RN2119}) and is still considered as a fundamental problem in constructing high-reliable systems today \citep{RN4069}.
In the literature, the basic problem is to reach an agreement in a pre-synchronized message-passing system of $n$ distributed \emph{nodes} in which up to $f$ such \emph{nodes} can fail arbitrarily.
During the early days, the so-called \emph{nodes} were often interpreted as computing components such as processors and servers, while the communicating components were largely considered as passive mediums.
In the core abstraction, an initiator node (the \emph{General}) is expected to broadcast a message $m$ containing a single value $v$ in a synchronous round-based system.
All \emph{correct} nodes in the network can receive this message in the same communication round providing that the \emph{General} is correct.
But the \emph{General} cannot always be correct.
And a \emph{faulty} (i.e., being not correct) \emph{General} can send arbitrary messages that contain inconsistent but still valid values toward different receivers (the \emph{Lieutenants}).
So it remains for these \emph{Lieutenants} to exchange their messages and to agree upon the same value in a bounded number of the following communication rounds.
Needless to say, an amount of \emph{Lieutenants} can also be faulty and exchange arbitrarily inconsistent messages in the network.
To exclude trivial solutions such that all correct \emph{Lieutenants} always agree upon a constant value, a correct \emph{Lieutenant} should agree upon the value $v$ contained in the broadcasted message $m$ whenever the \emph{General} is correct.

For decades, great efforts have been devoted to solving this problem.
In \cite{RN2119}, interactive consensus algorithms with oral and authenticated messages are provided.
The oral-message algorithm is first presented under fully connected networks and can then be extended to $3f$-regular networks.
The authenticated-message algorithm can be extended to arbitrarily connected networks but relies on authentication protocols.
However, a high-reliable authentication protocol might be impractical in real-world real-time systems.
And in considering efficiency, the oral-message algorithm should recursively call the sub-algorithms up to $(n-1)\cdots(n-f)$ times, during which at least $\Omega(n^f)$ messages should be exchanged between correct \emph{Lieutenants}.
Such disadvantages greatly prevent these algorithms from practical applications.
From that on, attention is drawn to optimize the required communication rounds, message and computation complexity, etc.
For example, a hybrid EIG-based solution with shift operations makes a trade-off between resilience, complexity, and required rounds in \cite{RN4250}.
And the early-stopping algorithm in \cite{Dolev1990Early} can reduce the actual communication rounds to the low-bound $\min\{f_{r}+2,f+1\}$ providing that $f$ is on the order of $\sqrt{n}$ and $f_{r}\leqslant f$, where $f_{r}$ is the number of actual faulty nodes instead of the maximal allowed number $f$.
In \cite{Berman1970Optimal}, early-stopping algorithm with optimal rounds and optimal resilience $n\geqslant 3f+1$ is provided.
And more efficient EIG-based algorithms \citep{Garay2006Fully,Kowalski2013Synchronous,RN4148} are provided with polynomial message complexity.
For further reducing message complexity, a single-bit message protocol is provided in \cite{BPG1989} with $n>4f$.
And in \cite{Berman1989Towards}, optimal resilience is achieved with 2-bits messages at the expense of tripled communication phases with a minimal number of phase-kings.
Meanwhile, in \cite{SrikanthSimulating}, the authenticated broadcast can be simulated by unauthenticated broadcast primitive, with which a bunch of authenticated BA algorithms can be converted into corresponding unauthenticated ones in fully connected networks.
This primitive facilitates building much easier-understood BA algorithms like \cite{Toueg1987Fast} comparing with some early explorations like \cite{RN4136}.
For other important improvements for solving the classical problem, we refer to \cite{Garay2006Fully,Kowalski2013Synchronous}.

\subsection{From theory to reality}
During the time, some variants of the original problem are also introduced in satisfying specific requirements raised from specific practical perspectives.
For example, the required agreement can be approximate \citep{RN4093}, randomized \citep{RN1569}, or differential \citep{Garay2003}.
The communication network can be asynchronous \citep{Bracha1984,Bracha1985,Abraham2020}, partially connected \citep{Dolev1981Unanimity}, or even sparse \citep{Dwork1986,Upfal1992}.
And the BFT algorithms can be self-stabilizing \citep{RN4099,Daliot2009Self}, self-optimizing \citep{Schmid2018}, just to name a few.
On one side, some variants, such as the approximate agreements and the randomized agreements, are often easier than the original problem.
But the approximate agreements (convex \citep{VectorConsensus2013,MultidimensionalAA2013} or not \citep{RN4093}) cannot reach exact $(0,1)$-agreement with just the fault-tolerant averaging functions \citep{RN3888}.
The randomized solutions \citep{King2011Breaking,RN4356} are provided with the assumption of some weaker adversary who cannot always know the pseudo-random numbers generated in the system.
This assumption is quite doubtful in the context of safety-critical systems whose failure rates are often required to be much better than $10^{-9}$ \citep{RN4551,RN965,Kopetz2004Hypothesis,RN2879}.
On the other side, the differential, asynchronous, self-stabilizing BA and the classical BA upon partially connected networks are all no easier than the original problem.
In a sense, all the harder BA problems can be handled with first solving the classical BA problem and then extending the solutions to those advanced problems.

However, up to now, most efficient deterministic synchronous BA algorithms (with early-stopping \citep{RN4148} or not \citep{Kowalski2013Synchronous}) are only presented upon fully connected reliable communication networks.
It remains for real-world systems to obtain adequate resources to satisfy these communication requirements.
Unfortunately, real-world communicating systems can hardly be both reliable and convenient in providing high network connectivity.
For example, bus-based networks can conveniently simulate fully connected networks of computing components, but they naturally lack in fault-containment of the communicating components and thus are often with low reliability.
Although switch-based networks can provide fault-containment of communicating links, it is still hard to support high connectivity among communicating components.
Consequently, practical BFT mechanisms in real-world systems, even the ones in the aerospace standards such as time-triggered protocol \citep{as6003} and TTEthernet \citep{as6802}, take some local \emph{filtering} schemes \citep{RN4551} to convert Byzantine faults of some communicating components into non-Byzantine ones.
As the faults of these local \emph{filters} (also referred to as the guardians or the monitors) cannot be tolerated in algorithms, these \emph{filters} together with the corresponding communicating components must be implemented and verified very carefully in practice to show adequate assumption coverage.

There are also BA solutions that tolerate malign faults in both computing components and communicating components, but the complexity in computation and communication is often high.
In \cite{Yan1989Achieving}, $\lfloor (n-1)/3 \rfloor$ computing and  $\lfloor n/2 \rfloor-1$ communicating Byzantine faults can be tolerated in $f+2$ communication rounds under fully connected network with $n$ computing components.
In \cite{WangByzantine}, by integrating broadcast network and fully connected network into fully connected multi-group networks (called $\mu$-$\gamma$ FCN), the agreement can be reached with fewer node-degrees and communication rounds than those required in former broadcast networks and fully connected networks.
And this EIG-based solution can also tolerate malign faults in both processor-groups and transmission mediums.
However, as transmission mediums are still considered as passive components, algorithms are designed only for processors, which restricts the efficiency and resilience of the agreement.
For example, although the local majority operation (referred to as \emph{LMAJ} in \cite{WangByzantine}) can optimize required rounds and message complexity, a dynamic message storage tree with $O(\gamma^2)$ entries should be maintained in each computing component, where $\gamma$ is proportional to $n$ when $\mu$ is fixed.
Besides, all $\gamma$ groups need to be fully connected by $\gamma(\gamma-1)/2$ communicating components in which at most $\lfloor \gamma/2 \rfloor-1$ are allowed to be faulty.
And as a faulty group can be generated by $\lceil \mu/2 \rceil$ processors and at most $\lfloor (n/\mu-1)/3 \rfloor$ faulty groups can be tolerated, best resilience can only be achieved in considering best-cases.
For worst-cases, $\lceil (\sqrt{n}+2)^2/6 \rceil$ Byzantine processors is enough to break down the system.
Now as these specific BA solutions are built upon special hybrid networking technologies (which might be not applicable to most distributed systems), the overall assumption coverage gained in these hybrid networks is not clear.
For the hard limitations and some new advances in hybrid-networking-based BA solutions, we refer to \cite{Khan2019}.

\subsection{From complete solutions to incomplete solutions}
Over the years, there are also many other fault-tolerant solutions built upon partially connected networks.
For example in \cite{RN3844}, BFT synchronization of hypercube networks is presented.
It shows that $m$ Byzantine faults can be tolerated in synchronizing hypercube network with connectivity greater than $\max(2m+1,3m-2)$.
This solution is proposed to synchronize the communication network in a parallel computer system designed as a scalable hypercube, which also demands expensive network connectivity.
From a practical perspective, as the sizes of the distributed systems are continually growing while the number of connections in any local component is severely bounded, for maintaining the required failure rates of the components being constant, a small number of the correct local components (the \emph{poor} ones) should be allowed to be surrounded (or overwhelmed) by the faulty components and thus at worst to be equivalently regarded as the faulty components \citep{Dwork1986}.

\cite{Dwork1986} shows that to extend BFT protocols such as BA to bounded-degree networks, some secure communication (\emph{communication} for short) protocols upon the bounded-degree networks can be built first, with which the fault-tolerant protocols originally provided upon fully connected networks can be simulated to some extent afterwards.
But to design such a communication protocol upon sublinear-degree networks, the expense is that some correct nodes might be connected to an overwhelming number of faulty nodes.
Namely, as the node-degree $d$ is required to be sublinear to $f$ while the failure rates of the components are required to be fixed, we can easily construct some cases in which some of the correct nodes are surrounded by faulty ones.
In this situation, only \emph{incomplete} \citep{BPG1989} protocols can be desired.
In \cite{BPG1989}, some \emph{incomplete} BFT protocols are also proposed in partially connected networks by employing several basic primitives.
In \cite{Upfal1992}, a linear number of Byzantine faults can be tolerated in a.e. BFT protocols upon some constant-degree expanders.
However, high computational complexity is demanded in \cite{Upfal1992}.
In \cite{Chandran2010}, the computational complexity is reduced by allowing the node-degree to be polylogarithmic (not \emph{constant}, but can still be viewed as \emph{sparse}).
However, the construction of the communication networks and the multi-layer fault-tolerant protocols \citep{Garay2003} are complex and with high computation and message complexities.
In \cite{Jayanti2020}, it shows that more efficient transmission schemes exist upon sparsely connected communication networks.
However, the construction of such networks is not explicit.
Also, these solutions mainly aim for reaching secure communication between the so-called \emph{privileged} nodes.
In constructing upper-layer BFT protocol like BA, the required complexity and execution time of the secure-communication-based BA are at least polynomial to those of the underlying secure communication protocols.
To break this barrier, some probabilistic BFT solutions are investigated \citep{BENOR1996329boundeddegree,King2006,King2011Breaking} at the expense of some additional possibilities of system failures.

Nevertheless, in the desired \emph{incomplete} protocols, although some correct nodes might be inevitably \emph{poor} \citep{Dwork1986} (or say being \emph{given up} \citep{Chandran2010}, i.e., being connected to an overwhelming number of faulty nodes) in bounded-degree networks, the number of such \emph{poor} nodes can be strictly bounded.
In this manner, desired deterministic BFT communication protocols can be provided, in which the faulty nodes together with the \emph{poor} ones can be regarded as the faulty nodes in an equivalent system with relatively higher failure rates of the nodes (i.e., with a bounded $\mu>1$).
Further, to give an asymptotic bound of the proportion of poor nodes, by denoting the maximal number of the \emph{poor} nodes in the $(\alpha,\mu)$-resilient \emph{incomplete} protocol as $\epsilon=(\mu-1)\alpha n$, when $\lim_{n\to \infty}(\epsilon/(n-f))=\lim_{n\to \infty}(\mu-1)\alpha/(1-\alpha)=0$, the protocol is said to be an \emph{almost everywhere} (a.e.) $\epsilon$-\emph{incomplete} protocol \citep{Dwork1986,BPG1989}.
It should be noted that in this definition, $\alpha$ is allowed to be a function $\alpha(n)=n^{-\epsilon_1}$ for some constant $\epsilon_1>0$ with $\epsilon_1\to 0$ as $d\to n-1$ (see \cite{Dwork1986}).
So an $(\alpha,\mu)$-resilient $\epsilon$-\emph{incomplete} protocol is a.e. $\epsilon$-\emph{incomplete} if $\lim_{n\to \infty}\mu\alpha=0$ in this context.
And obviously, if $\mu$ is a constant being independent of $f$ and $n$ \citep{Upfal1992}, the protocol is an a.e. protocol.
From the practical perspective, it often suffices to make $\mu$ being an constant, with which an incomplete BFT system upon some low-degree network can be viewed as a complete BFT system upon fully connected network where the components are allowed with a relatively higher but still constant failure rate.

However, there are several problems with the incomplete communication protocols.
Firstly, these protocols are mainly designed for the secure communication between the non-poor-correct (i.e., correct and not poor, \emph{npc} for short) nodes.
A faulty or poor node can still make inconsistent communication with the npc nodes.
Secondly, it often requires several basic rounds (at least $O(\log n)$, see \cite{Upfal1992,UPFAL1994312,Chandran2010,Jayanti2020}) to simulate an upper-layer communication-round between two npc nodes, which is a factor of the overall required basic rounds for the upper-layer protocol such as BA.
Thirdly, in the case of the partially connected networks, to select the correct values among the faulty ones in all messages from all corresponding paths, the required fault-tolerant process might be with high complexity.
For example, in \cite{Upfal1992}, each correct node needs to select a subset of the messages coming from all the paths by referencing all possible combinations of the faulty nodes, with which exponential complexity is demanded.
Although this complexity is reduced in \cite{Chandran2010} to polynomial, it relies on a multi-layer transmission scheme upon a complicated polylogarithmic-degree network.
And in all these solutions, each correct node should know the topology of the network, which is unscalable and also excludes dynamical networks.

\subsection{Other related solutions}
In practice, many \emph{fault-tolerant} solutions work for partially connected networks do not achieve Byzantine resilience.
In \cite{RN4129}, fault-tolerance for complete bipartite networks are presented but only fail-silence faults are considered.
Some other fault-tolerant systems upon circulant graphs, meshes of trees, butterflies, and hypercubes \citep{Leighton1992On,BruckFault,Farrag2008Fault} also share with the similar assumption coverage.
In \cite{RN3858,RN3857}, real-world protocols are presented in ring topologies to tolerate \emph{as many malfunctions as possible}.
However, the reliability of these protocols is restricted by possible Byzantine behaviors.
Especially, it is often hard to show adequate assumption coverage for the real-world systems with various active communicating components such as the TSN (time-sensitive-networking) switches \citep{tsn}, OpenFlow switches \citep{LaraNetwork}, customized switches, etc.

In \cite{TDWALDEN}, an efficient self-stabilizing BFT synchronization solution is proposed upon small scale bipartite networks.
However, the expected $O(f^f)$ stabilization time would be prohibitively high with the increasing of $f$.
In reducing the stabilization time, some self-stabilizing BA primitives can be good alternatives.
But the self-stabilizing BA primitives \citep{Daliot2009Self, RN4099, daliot2006self} are mainly proposed for systems with fully connected networks.
Practically, as computation and communication are often modularized into unreliable \emph{computing components} and \emph{communicating components} in the systems, it is interesting to construct efficient BA upon bipartite networks.
In \cite{ASBBDNFTP}, the classical broadcast primitive is extended to some non-bipartite sparse networks with fault-tolerant propagation.
However, the results of fault-tolerant propagation cannot directly extended to bipartite networks.
Meanwhile, the extended broadcast primitive is still built upon localized secure communication protocols.
In further avoiding secure communication and the related protocols, \cite{BPLSNHSAC} proposes that the classical Byzantine adversary can be extended to the multi-scale Byzantine adversaries with reinvestigating the practical system assumption coverage.

\section{The system and the problems}
\label{sec:Model}
\subsection{System model}
The system $\mathcal{S}$ discussed in this paper consists of $n$ distributed components, denoted as the node set $V=\{1,2,\dots,n\}$.
Following our previous work \cite{ASBBDNFTP}, we assume that up to $f=\alpha n$ nodes in $V$ can fail arbitrarily and all the other nodes are correct, where $\alpha\in [0,1)$ is called the Byzantine resilience with the trivial rounding problem being ignored.
The point-to-point bidirectional communication channels are represented as the undirected edge set $E$, with which the undirected graph $G=(V,E)$ forms.
Specifically, in the bipartite cases, the discussed bipartite network is represented as the undirected graph $G_{bi}=( V_A \cup V_B,E_{bi} )$, where the disjoint node-subsets $V_A$ and $V_B$ can be interpreted as any kinds of distributed components, such as the end systems (for computation) and customized switches (for communication) in Ethernet.
The numbers of nodes in $V_A$ and $V_B$ are respectively denoted as $n_A=|V_A|$ and $n_B=|V_B|$, i.e., $V=V_A \cup V_B$ includes $n=n_A+n_B$ distinct nodes.
The sets of all correct nodes in $V_A$ and $V_B$ are respectively denoted as $U_A$ and $U_B$, by which we also denote $U=U_A\cup U_B$.
For our interest, there can be up to $f_A=\alpha_A n_A$ and $f_B=\alpha_B n_B$ Byzantine nodes $T_A$ and $T_B$ in $V_A$ and $V_B$, respectively, where $\alpha_A\in [0,1)$ and $\alpha_B\in [0,1)$ can be independently set according to the respective failure rates of real-world components.
As the failures of the communication channels can be equivalent to the failures of the distributed components (for example, the nodes in each side of the bipartite networks), the edges in $G$ are assumed reliable.

In concrete cases, the system $\mathcal{S}$ can be further interpreted as the specific systems, such as the broadcast system and the agreement system (defined shortly after).
For example, the specific case $\mathcal{S}$ being a broadcast system is handled in \cite{ASBBDNFTP}.
In \cite{ASBBDNFTP}, we have assumed that the system is executed in synchronous rounds (\emph{round} for short), the adversary is \emph{strong}, and the set $T$ cannot be changed during the execution of $\mathcal{S}$.
These basic assumptions and the basic system model proposed in \cite{ASBBDNFTP} are still valid here.
Nevertheless, there are also some differences.
Firstly, as the computing and communicating components are distinguished as two different kinds of nodes in $G_{bi}$, each synchronous round is composed of two semi-rounds (or saying \emph{phases}).
Namely, during the first phase of each round of the execution of $\mathcal{S}$, every correct node $i\in U_A$ can synchronously send its message to all its neighbours $N_i\subseteq V_B$.
Then, during the second phase of the same round, every correct node $j\in U_B$ would send its message according to the messages it received during the first phase.
With some BFT operations, $j$ can decide if (and when) its message needs to be generated and sent.
Secondly,  we assume that the algorithms provided for the nodes are locally identical in the range of $U_A$ and $U_B$.
Namely, the algorithms are uniform in each side of the bipartite network.
The algorithms for any two nodes $i\in U_A$ and $j\in U_B$ can be different.
Thirdly, unlike the $1$-bit messages employed in the broadcast systems, multi-value messages are also allowed in the agreement systems.

\subsection{The basic problems}
In the classical BA problem \citep{PSL1980}, the system $\mathcal{S}$ is required to reach the desired agreement for the \emph{General} (can be faulty or correct) in some bounded rounds.
As the original network $G$ is assumed as $K_n$ (the $n$-nodes complete graph), the \emph{General} can be assumed as any node in $V$ or an external system.
But in the partially connected networks, as there might be no node with a sufficient node-degree, we always assume an external \emph{General} who logically initiate the corresponding \emph{Lieutenants} (with an initial value for each \emph{Lieutenant}) before the beginning of the first round in every execution of $\mathcal{S}$ in the BA problem.
And for reaching the desired agreement for this \emph{General} upon partially connected networks, these \emph{Lieutenants} begin to exchange messages during the execution of $\mathcal{S}$.
In the classical solution \citep{RN2119}, these \emph{Lieutenants} recursively set themselves as the sub-rank \emph{Generals} and respectively initiate the corresponding sub-rank \emph{Lieutenants} in each round.
But in the partially connected networks, this can also hardly do with the insufficient node-degrees.
As an alternative, following \cite{SrikanthSimulating}, here we intend to simulate the relay-based broadcast upon the partially connected networks for the top-rank \emph{Lieutenants} (or saying the second-rank \emph{Generals}) in a nonrecursive way.
With the prior works \citep{SrikanthSimulating,Toueg1987Fast,RN4346}, it is easy to see that if the relay-based broadcast can be simulated upon some partially connected networks to some extent, the corresponding BA solution follows.
Further, by allowing a small number of correct nodes to be regarded as the faulty ones (the \emph{poor} nodes \citep{Dwork1986}), an agreement among most of the correct nodes can be reached in some bounded-degree networks (which is of practical significance in real-world large-scale systems), providing that the corresponding incomplete broadcast can be simulated efficiently in such networks.

Here we define the related systems.
Following \cite{ASBBDNFTP}, $\mathcal{S}$ is a broadcast system upon $G$ iff $\mathcal{S}$ can simulate the authenticated broadcast like \cite{SrikanthSimulating} upon $G$.
$\mathcal{S}$ is an agreement system upon $G$ iff $\mathcal{S}$ can reach the required BA \citep{RN2119} upon $G$.
In generally measuring the Byzantine resilience, $\mathcal{S}$ is an $\alpha$-resilient $\mathtt{X}$ system upon $G$ iff $\mathcal{S}$ can completely reach the requirement of the $\mathtt{X}$ system upon $G$ in the presence of $\alpha n$ Byzantine nodes, where $\mathtt{X}$ can be broadcast, agreement, etc.
And $\mathcal{S}$ is an $(\alpha,\mu)$-resilient incomplete $\mathtt{X}$ system upon $G$ with $1<\mu<\alpha^{-1}$ iff $\mathcal{S}$ can reach the requirement of the $\mathtt{X}$ system in at least $(1-\mu\alpha) n$ nodes upon $G$.
For completeness, the $\alpha$-resilient systems are also the $(\alpha,\mu)$-resilient systems with $\mu=1$.
Again, we consider only the deterministic worst-case solutions.

With these, the problems discussed in this paper are to provide the corresponding deterministic BFT systems upon $G\neq K_n$.
For our specific interest here, the systems upon the bipartite graph $G_{bi}=( V_A \cup V_B,E_{bi} )$ are specially considered, as the two sides of $G_{bi}$ may well correspond to some real-world computing components and communicating components, with which some extra efficiency can be gained upon specific underlying communication networks.

\section{A review of the broadcast systems upon $K_n$}
\label{sec:Broadcast_1}
The broadcast system upon $K_n$ is first provided in \cite{SrikanthSimulating} to simulate the authenticated broadcast in synchronous peer-to-peer networks.
With this, agreement systems upon $K_n$ are provided by employing the broadcast system to implement the so-called \emph{broadcast primitive} \citep{Toueg1987Fast}.
It is interesting to ask to what extent this primitive can be extended to partially connected networks.
In \cite{ASBBDNFTP}, the broadcast primitive upon $K_n$ has been extended to some non-bipartite bounded-degree networks with a simple system structure.
Here our goal is to extend the broadcast primitive to bipartite networks.
So, before directly handling the bipartite cases, we first shortly introduce the system structure proposed in \cite{ASBBDNFTP}.
Then, by reviewing the broadcast primitive upon $K_n$ with the proposed system structure, we show that the broadcast systems can be constructed under a basic strategy which can be extended to the broadcast systems upon arbitrarily connected networks.
If not specified, the broadcast system mentioned in this section is upon $K_n$.

\subsection{The general broadcast system}
Following \cite{ASBBDNFTP}, we can represent the general broadcast system $\mathcal{D}$ as
\label{subsec:generalD}
\begin{eqnarray}
\label{eq:functions_transfer_general}&&\vec{x}(k)=D_x(\hat{\mathbf{X}}(k-1))+D_u(\vec{u}(k)) \\
\label{eq:functions_estimate_general}&&\hat{\mathbf{X}}(k)=(\mathbf{I}+\mathbf{A})\odot ({\vec 1}^T \otimes \vec{x}(k)+\mathbf{F}(k))\\
\label{eq:functions_yield_general}&&\vec{y}(k)=D_y(\hat{\mathbf{X}}(k))
\end{eqnarray}
where the vector $\vec{x}(k)\in \mathbb V^{n}=\{0,1\}^{n}$ is the system state (at the discrete time $k$), the matrix $\hat{\mathbf{X}}(k)=[ \hat{\vec{x}}^{(1)}(k),\dots,\hat{\vec{x}}^{(n)}(k)]\in \mathbb V^{n\times n}$ is the estimated system state in all nodes (each column for a distinct node), and the vectors $\vec{y}(k)\in \mathbb V^{n}$ and $\vec{u}(k)\in \mathbb V^{n}$ are respectively the decision vector and the input vector of the broadcast system.
The noise matrix $\mathbf{F}(k)\in \mathbb F^{n\times n}=\{0,-1,1\}^{n\times n}$ has up to $f$ nonzero (being arbitrarily valued) rows, denoted as $\mathbf{F}(k)\in \Upsilon^{[f]}$.
The adjacency matrix of the arbitrarily connected network $G$ is denoted as $\mathbf{A}$.
The identity matrix $\mathbf{I}$ is with the same size of $\mathbf{A}$.
The operator $\otimes$ computes the Kronecker product of two matrices.
The operator $\odot $ computes a matrix with elements $x_{i,j}=y_{i,j}z_{i,j}$ for the same-sized matrices, where $x_{i,j}$, $y_{i,j}$, and $z_{i,j}$ denote the elements in the $i$th row and $j$th column of the matrices.
The set of all possible executions of $\mathcal{D}$ is denoted as $\Lambda_{D}$.
An execution $\chi\in \Lambda_{D}$ is referred to as an $f$-Byzantine execution, denoted as $\chi\in \Lambda_{D}^{[f]}$, iff all noise matrices in this execution are in $\Upsilon^{[f]}$.
A function $s:\mathbb Z\to \mathbb V$ is also called a signal.
An $n$-dimensional signal $\vec{s}:\mathbb Z\to \mathbb V^n$ in a concrete execution $\chi\in \Lambda_{D}$ is denoted as $\vec{s}|_\chi$.

For convenience, we use the $0$-norm $\|\vec{q}\|$ to denote the number of the nonzero elements in the vector $\vec{q}$.
Meanwhile, the discrete Dirac signal $\delta$, the discrete Heaviside step signal $H$, the absolute-value signal $|s|$, the shift signal $s_{+k_0}$, and the other operators on the signals are all defined in the way of \cite{ASBBDNFTP}.
With the bounded $k_H$ and $k_\delta$, an $(\alpha,\mu)$-resilient $\mathcal{D}$ is with the $k_H$-\emph{Heaviside} property iff
\begin{eqnarray}
\label{eq:Heaviside_ex}
(\forall i,j\in V_0 :u_i\equiv u_j)\to\nonumber \\
(\exists P\subseteq U: |P|\geqslant (1-\mu\alpha)n \land \forall i\in P : \nonumber \\
\exists 0\leqslant k_1<k_H : y_i=\sum u_{+k_1})
\end{eqnarray}
holds for every $\chi\in \Lambda_{D}^{[\alpha n]}$, where $V_0$ is the set of the initialized (by the external \emph{General}) nodes and $\sum s$ is the integral signal of the $s$ signal.
Meanwhile, $\mathcal{D}$ is with the $k_\delta$-\emph{Dirac} property iff
\begin{eqnarray}
\label{eq:Dirac_ex}
\exists P\subseteq U, k_1\leqslant\dots\leqslant k_m< k_1+k_\delta: |P|\geqslant (1-\mu\alpha)n \land  \nonumber \\
 \forall i,j\in P:| y_i-y_j |\leqslant \sum_{r=1}^{m}\delta_{+k_r}
\end{eqnarray}
holds for every $\chi\in \Lambda_{D}^{[\alpha n]}$.
All these definitions are just the same as \cite{ASBBDNFTP}.

\subsection{The Heaviside constraints}

Firstly, we look to some basic constraints imposed by the $1$-Heaviside property in the broadcast system without any faulty local system.

\begin{lemma}
\label{lemma_Dx0Dy1}
If $\mathcal{D}$ satisfies the $1$-Heaviside property, then $D_y(\vec1)=1$ and $D_x(\vec0)=0$.
\end{lemma}
\begin{IEEEproof}
Firstly, there exists $\chi_1\in \Lambda_{D}^{[0]}$ which makes ${\vec{u}}|_{\chi_1}(0)=\vec{1}$ and $\mathbf{F}_{\chi_1}(0)=0$.
By the requirement of the $1$-Heaviside property, $\vec{y}|_{\chi_1}(0)=\vec1$ should hold.
With (\ref{eq:functions_yield_general}), (\ref{eq:functions_estimate_general}) and (\ref{eq:functions_transfer_general}), $\vec{y}|_{\chi_1}(0)=D_y(\hat{\mathbf{X}}|_{\chi_1}(0))=D_y({\vec 1}^T \otimes {\vec{x}}|_{\chi_1}(0))=D_y({\vec 1}^T \otimes \vec{1})$.
Thus $D_y(\vec1)=1$ holds.

Secondly, there exists $\chi_2\in \Lambda_{D}^{[0]}$ which makes ${\vec{u}}|_{\chi_2}(k)=\vec{0}$ and $\mathbf{F}_{\chi_2}(k)=0$ for all $k\geqslant 0$.
By the requirement of the $1$-Heaviside property, $D_y(D_x(\hat{\mathbf{X}}|_{\chi_2}(k-1)))=0$ should hold for every $k\geqslant 0$.
As $D_y(\vec1)=1$, we have $D_x(\hat{\vec{x}}^{(i)}|_{\chi_2}(k-1))=D_x({\vec{x}}|_{\chi_2}(k-1))\neq 1$ for every $i\in V$ when $k>0$, i.e., $D_x(\hat{\vec{x}}^{(i)}|_{\chi_2}(k-1))=D_x({\vec{x}}|_{\chi_2}(k-1))=0$.
With (\ref{eq:functions_transfer_general}), we get $\vec{x}|_{\chi_2}(k)=\vec0$ for all $k> 0$.
Now as $D_x({\vec{x}}|_{\chi_2}(1))=0$, $D_x(\vec0)=0$ holds.

\end{IEEEproof}

\begin{lemma}
\label{lemma_Dx00}
If $\mathcal{D}$ satisfies the $1$-Heaviside property and ${\vec{u}}(k)=\vec{0}$ holds for all $k\leqslant k_0$ where $k_0\in \mathbb N$, then $D_x(\hat{\mathbf{X}}(k-1))=\vec0$ and $\vec{x}(k)=\vec0$ hold if $0\leqslant k\leqslant k_0$.
\end{lemma}
\begin{IEEEproof}
By the second part of the proof of Lemma \ref{lemma_Dx0Dy1}, when ${\vec{u}}|_{\chi_2}(k)=\vec{0}$ and $\mathbf{F}_{\chi_2}(k)=0$ for all $k\leqslant k_0$, we have $D_x(\hat{\vec{x}}^{(i)}|_{\chi_2}(k-1))=0$ when $0<k\leqslant k_0$.
Thus, when $0<k\leqslant k_0$, $D_x(\hat{\mathbf{X}}(k-1))=\vec0$ holds if $\forall k\leqslant k_0:{\vec{u}}(k)=\vec{0}$.

When $k=0$, as all $D^{(i)}$ are uniform in $\mathcal{D}$, $\hat{\vec{x}}^{(i)}|_{\chi_2}(-1)=\hat{\vec{x}}^{(j)}|_{\chi_2}(-1)$ should hold for every $i,j\in V$.
As $D_x(\hat{\mathbf{X}}|_{\chi_2}(-1))\neq \vec1$, we have $D_x(\hat{\mathbf{X}}(-1))= \vec0$ if ${\vec{u}}(0)=\vec{0}$.
Now as $D_x(\hat{\mathbf{X}}(k-1))=\vec0$ holds for all $k$ satisfying $0\leqslant k\leqslant k_0$, with (\ref{eq:functions_transfer_general}) we also have $\vec{x}(k)=\vec0$.

\end{IEEEproof}

\begin{corollary}
\label{co_Assume_0}
If $\mathcal{D}$ satisfies the $1$-Heaviside property, then we can always assume $D_x(\hat{\mathbf{X}}(-1))=\vec0$.
\end{corollary}
\begin{IEEEproof}
By the second part of proof of Lemma \ref{lemma_Dx00}, as all $D^{(i)}$ are uniform in $\mathcal{D}$, it means $D_x(\hat{\vec{x}}^{(i)}(-1))=0$ whenever $u_i(0)=0$.
And when $u_i(0)=1$, with (\ref{eq:functions_transfer_general}) we have $x_i(0)=1$ in regardless of the value of $D_x(\hat{\vec{x}}^{(i)}(-1))$.
Thus $D_x(\hat{\mathbf{X}}(-1))=\vec0$ can always be assumed.

\end{IEEEproof}

Thus, we assume $D_x(\hat{\mathbf{X}}(-1))=\vec0$ in the rest of the paper.
Now, we show constraints imposed by the $1$-Heaviside property in considering at most one faulty local system.
\begin{lemma}
\label{lemma_restriction_Dy}
If $\mathcal{D}$ satisfies the $1$-Heaviside property, then
$\forall \vec{q}\in \mathbb V^n: \exists \chi \in \Lambda_{D}^{[1]}: D_y(\hat{\mathbf{X}}|_{\chi}(0))=\vec{q}$.
\end{lemma}
\begin{IEEEproof}
According to (\ref{eq:functions_estimate_general}), the state-estimation matrix $\hat{\mathbf{X}}(k)=\vec{1}^T\otimes(D_x(\hat{\mathbf{X}}(k-1))+{\vec{u}}(k))+\mathbf{F}(k)$ is determined by $\vec{x}(k)$, ${\vec{u}}(k)$ and $\mathbf{F}(k)$.
By definition, $D_x(\hat{\mathbf{X}}(-1))$ can be $\vec{0}$, ${\vec{u}}(0)$ can take arbitrary values in $\mathbb V^n$, and $\mathbf{F}(0)$ can have one arbitrary Byzantine row.
Thus for every $\vec{p}\in \mathbb V^n$, $\vec{r}\in \mathbb F^n$, and $r\in V$, there exists an execution $\chi \in \Lambda_{D}^{[1]}$ in which $\hat{\mathbf{X}}|_{\chi}(0)=\vec{1}^T\otimes \vec{p}+\vec{r}^T\otimes\vec{e}_r$ holds, where $\vec{e}_r$ denotes the $r$th column vector of the identity matrix $I$.

If there exists $\vec{p}_0,\vec{p}_1\in \mathbb V^n$ with which $D_y(\vec{p}_0)=0 \land D_y(\vec{p}_1)=1$ and $\exists r_0\in V: \vec{p}_1-\vec{p}_0 =\vec{e}_{r_0}$ holds, we can always choose an execution $\chi_0 \in \Lambda_{D}^{[1]}$ in which $\hat{\mathbf{X}}|_{\chi_0}(0)=\vec{1}^T\otimes \vec{p}_0+\vec{{r}}^T\otimes\vec{e}_{r_0}$ holds.
So we have $\hat{\mathbf{X}}|_{\chi_0}(0)=(\vec{1}-\vec{{r}})^T\otimes \vec{p}_0+\vec{{r}}^T\otimes\vec{p}_1$.
Since $D_y((\vec{1}-\vec{{r}})^T\otimes \vec{p}_0+\vec{{r}}^T\otimes\vec{p}_1)=\vec{{r}}$ where $\vec{{r}}\in \mathbb F^n$ is arbitrarily valued, we would arrive the conclusion if such $\vec{p}_0$ and $\vec{p}_1$ exist.

Suppose no $\vec{p}\in \mathbb V^n$ satisfies $D_y(\vec{p})=0 \land D_y(\vec{p}+\vec{e}_{r})=1$ or $D_y(\vec{p})=1 \land D_y(\vec{p}+\vec{e}_{r})=0$ for any $r\in V$.
Denoting $D_y(\vec{0})=v\in \mathbb V$, we have $\forall r_1,r_2,\dots,r_n\in V:D_y(\vec{e}_{r_1})=v \land D_y(\vec{e}_{r_1}+\vec{e}_{r_2})=v \land \dots \land D_y(\sum_{r=1}^{n}\vec{e}_{r})=v$.
Thus $\forall \vec{p}\in \mathbb V^n:D_y(\vec{p})=v$.
A contradiction with the $1$-Heaviside property.

\end{IEEEproof}

Then, in considering up to $f$ faulty local systems, we have the following observations.
\begin{lemma}
\label{lemma_restriction_DyDx}
$\mathcal{D}$ satisfies the $1$-Heaviside property in the presence of $\Upsilon^{[f]}$ iff
$\forall \vec{q}\in \mathbb V^n: (D_x(\vec{q})=D_y(\vec{q})=0 \lor {\lVert \vec{q}\rVert}>f) \land (D_x(\vec{q})=D_y(\vec{q})=1 \lor {\lVert \vec{q}-\vec1\rVert}>f)$.
\end{lemma}
\begin{IEEEproof}
Firstly, assume $\forall \vec{q}\in \mathbb V^n: (D_x(\vec{q})=D_y(\vec{q})=0 \lor {\lVert \vec{q}\rVert}>f) \land (D_x(\vec{q})=D_y(\vec{q})=1 \lor {\lVert \vec{q}-\vec1\rVert}>f)$.
When ${\vec{u}}(k)=\vec{0}$ holds for all $k\leqslant k_0$ where $k_0\in \mathbb N$, with (\ref{eq:functions_transfer_general}), (\ref{eq:functions_estimate_general}) and (\ref{eq:functions_yield_general}), we get ${\vec{x}}(k)=\vec0$ and thus ${\vec{y}}(k)=\vec0$ for all $k\leqslant k_0$.
And when ${\vec{u}}(k)=\delta[k-k_0]$, we get ${\vec{x}}(k_1)=\vec1$ and thus ${\vec{y}}(k_1)=\vec1$ for all $k_1\geqslant k_0$.
Thus, $\mathcal{D}$ satisfies the $1$-Heaviside property in the presence of $\Upsilon^{[f]}$.

Conversely, to satisfy the $1$-Heaviside property, if ${\vec{u}}|_{\chi}(0)=\vec0$ in an execution $\chi\in \Lambda_{D}^{[f]}$, ${\vec{y}}|_{\chi}(0)=\vec0$ should hold.
By (\ref{eq:functions_yield_general}) and (\ref{eq:functions_estimate_general}), it means $D_y(D_x(\hat{\mathbf{X}}|_{\chi}(-1))+\vec{\upsilon}^{(i)} |_{\chi}(0))=0$ should hold for any $\vec{\upsilon}^{(i)} |_{\chi}(0)$.
As there are up to $f$ faulty nodes and $D_x(\hat{\mathbf{X}}|_{\chi}(-1))$ can be valued as $\vec0$ in $\chi$, $D_y(\vec{q})=0$ should hold for every $\vec{q}\in \mathbb V^n$ that satisfies ${\lVert \vec{q}\rVert}\leqslant f$.
In similar ways, by setting ${\vec{u}}|_{\chi'}(0)=\vec1$, $D_y(\vec{q})=1$ holds for every $\vec{q}\in \mathbb V^n$ with ${\lVert \vec{q}-\vec1\rVert}\leqslant f$.

In any an execution $\chi\in \Lambda_{D}^{[f]}$, if ${\vec{u}}|_{\chi}(1)={\vec{u}}|_{\chi}(0)=\vec0$, according to the $1$-Heaviside property, ${\vec{y}}|_{\chi}(1)=\vec0$ should also hold.
By (\ref{eq:functions_yield_general}) and (\ref{eq:functions_estimate_general}), $D_y(\vec{x}|_{\chi}(1)+\vec{\upsilon}^{(i)} |_{\chi}(1))=\vec0$ should hold.
Now assume there exists a $\vec{p}\in \mathbb V^n$ satisfying ${\lVert \vec{p} \rVert}\leqslant f \land D_x(\vec{p})=1$.
Then there exists an execution $\chi_1\in \Lambda_{D}^{[f]}$ satisfying $\vec{x}|_{\chi_1}(0)=\vec0$ and $\mathbf{F}_{\chi_1}^{[f]}(0)=\vec{1}^T\otimes \vec{p}$.
By (\ref{eq:functions_transfer_general}), we have $\vec{x}|_{\chi_1}(1)=D_x(\hat{\mathbf{X}}|_{\chi_1}(0))=D_x(\vec{1}^T\otimes \vec{p})=\vec1$.
Now as ${\lVert \hat{\vec{x}}^{(i)}|_{\chi_1}(1) \rVert}={\lVert \vec{1}+\vec{\upsilon}^{(i)} |_{\chi}(1) \rVert}\geqslant {\lVert \vec{1}\rVert}-{\lVert \vec{\upsilon}^{(i)} |_{\chi}(1) \rVert}=n-f$ holds for all $i\in V$, we have $D_y(\hat{\mathbf{X}}|_{\chi_1}(1))=\vec1$.
A contradiction with ${\vec{y}}|_{\chi}(1)=\vec0$.
Similarly, by setting ${\vec{u}}|_{\chi'}(0)=\vec1$, $D_x(\vec{q})=1$ holds for every $\vec{q}\in \mathbb V^n$ with ${\lVert \vec{q}-\vec1\rVert}\leqslant f$.

\end{IEEEproof}

\subsection{An impossible property}
According to the requirement of relay-based broadcast, for every $\chi\in \Lambda_{D}^{[f]}$, the output decision vector ${\vec{y}}$ should satisfy both the $1$-Heaviside and $1$-Dirac properties.
If, however, the $1$-Heaviside and $0$-Dirac properties could be satisfied in $\mathcal{D}$, the requirement of the agreement problem would be directly satisfied.
Here we show that any $\mathcal{D}$ has the structure generalized in \ref{subsec:generalD} cannot satisfy both the $1$-Heaviside and $0$-Dirac properties in the presence of a faulty local system.

\begin{lemma}
\label{lemma_1_dirac_impossible}
No $\mathcal{D}$ can satisfy both the $1$-Heaviside and $0$-Dirac properties in the presence of a faulty local system.
\end{lemma}
\begin{IEEEproof}
Suppose some $\mathcal{D}$ satisfies $1$-Heaviside, $0$-Dirac properties with a faulty local system.
According to the $1$-Heaviside property, there exists $\chi_1\in \Lambda_{D}^{[0]}$ during which the \emph{General} is correct and for all $k\geqslant 0$, ${\vec{u}}|_{\chi_1}(k)=\delta[k-k_0]\cdot \vec{1}$ and $ {\vec{y}}|_{\chi_1}(k)=H[k-k_0]\cdot \vec{1}$ holds for some a fixed $k_0\geqslant 0$.
Thus by (\ref{eq:functions_yield_general}), $D_y(\hat{\mathbf{X}}|_{\chi_1}(k))=H[k-k_0]$ should hold for some $\hat{\mathbf{X}}|_{\chi_1}$.
With such an $\hat{\mathbf{X}}|_{\chi_1}$, $D_y(\hat{\vec{x}}^{(i)}|_{\chi_1}(k_0))=1$ and $D_y(\hat{\vec{x}}^{(i)}|_{\chi_1}(k_0-1))=0$ hold for any $i\in V$ under $\mathbf{F}_{\chi_1}^{[0]}(k_0)=0$.

Now consider any a $\chi'_1\in \Lambda_{D}^{[1]}$ which only differs from $\chi_1$ with the $i$th element in the $r$th ($r\in V$) row of $\mathbf{F}_{\chi_1}^{[0]}(k_0)$ in the $k_0$th round.
We denote this noise matrix in round $k_0$ of $\chi'_1$ as $\mathbf{F}_{\chi'_1}^{[1]}(k_0)$.
In $\chi'_1$, as $\hat{\vec{x}}^{(j)}|_{\chi'_1}(k_0)$ is the same as that in $\chi_1$ for any $j\neq i$, we have $D_y(\hat{\vec{x}}^{(j)}|_{\chi'_1}(k_0))=1$ and $D_y(\hat{\vec{x}}^{(j)}|_{\chi'_1}(k_0-1))=0$ hold for each $j\neq i$.

According to the $0$-Dirac property, for this $\chi'_1$, $ \forall k\geqslant 0: \Delta\vec{y}|_{\chi'_1}(k)=\delta[k-k_1]\cdot \vec{1}$ holds for some a finite $k_1\geqslant 0$, where $\Delta \vec{y}(k)=\vec{y}(k)-\vec{y}(k-1)$ is the backward-difference of ${\vec{y}}(k)$.
Take (\ref{eq:functions_yield_general}) into it, we have $D_y(\hat{\mathbf{X}}|_{\chi'_1}(k_0))-D_y(\hat{\mathbf{X}}|_{\chi'_1}(k_0-1))=\delta[k_0-k_1]\cdot  \vec{1}$  by taking $k$ as $k_0$.
Now if $k_0\neq k_1$, it can only be $D_y(\hat{\mathbf{X}}|_{\chi'_1}(k_0))=D_y(\hat{\mathbf{X}}|_{\chi'_1}(k_0-1))$, which contradicts with $D_y(\hat{\vec{x}}^{(j)}|_{\chi'_1}(k_0))=1$ and $D_y(\hat{\vec{x}}^{(j)}|_{\chi'_1}(k_0-1))=0$ for any $j\neq i$.
So it can only be $k_0=k_1$ and $D_y(\hat{\mathbf{X}}|_{\chi'_1}(k_0))-D_y(\hat{\mathbf{X}}|_{\chi'_1}(k_0-1))=\vec{1}$, where $D_y(\hat{\vec{x}}^{(i)}|_{\chi'_1}(k_0))=1$ should hold.
As the $i$th element in the $r$th row of $\mathbf{F}_{\chi'_1}^{[1]}(k_0)$ can take arbitrary values, we have $\forall c_r\in \mathbb F: D_y(\vec{x}|_{\chi'_1}(k_0)+c_r\cdot\vec{e}_r)=1$, where $\vec{e}_r$ denotes the $r$th column vector of the identity matrix $I$.
As $\mathcal{D}$ has uniform $D_y$ and $D_u$ blocks, this holds for all local systems in $\mathcal{D}$.

As $r$ can be arbitrarily selected when we choose $\chi'_1$, without loss of generality we take $r$ as $1$ in $\chi'_1$.
Now consider another $\chi_{r+1}\in \Lambda_{D}^{[0]}$ which only differs from $\chi_r$ by changing the input signal from $\vec{u}|_{\chi_r}(k_0)$ to $\vec{u}|_{\chi_{r+1}}(k_0)$.
This $\vec{u}|_{\chi_{r+1}}(k_0)$ comes from a faulty \emph{General} and makes ${\vec{u}}|_{\chi_{r+1}}(k_0) ={\vec{u}}|_{\chi_{r}}(k_0)+c_r\cdot\vec{e}_r$ being true where $c_r\cdot\vec{e}_r=\vec{\upsilon}^{(i)}|_{\chi'_{r}}(k_0)$ is the noise vector measured by node $i$ in the $k_0$th round of $\chi'_r$.
By the proof of Lemma \ref{lemma_Dx00}, we have $D_x(\hat{\mathbf{X}}|_{\chi_{r+1}}(k_0-1))=\vec0$.
So ${\vec{y}}|_{\chi_{r+1}}(k_0)=D_y({\vec{u}}|_{\chi_{r+1}}(k_0))=D_y(\hat{\mathbf{X}}|_{\chi_{r}}(k_0))=1$ and ${\vec{y}}|_{\chi_{r+1}}(k_0-1)={\vec{y}}|_{\chi_{r}}(k_0-1)=0$.

Now consider any $\chi'_{r+1}\in \Lambda_{D}^{[1]}$ which only differs from $\chi_{r+1}$ with the $i$th element in the $(r+1)$th row of $\mathbf{F}_{\chi_{r+1}}^{[0]}(k_0)$ in round $k_0$.
We have $\forall c_{r+1}\in \mathbb F: D_y({\vec{u}}|_{\chi_{r+1}}(k_0)+c_{r+1}\cdot\vec{e}_{r+1})=1$.
Iteratively, we have $\forall c_1,\dots,c_n\in \mathbb F: D_y({\vec{u}}|_{\chi_{1}}(k_0)+\sum _{r=1}^{n}c_r\cdot\vec{e}_r)=1$.
As $\{\sum _{r=1}^{n}c_r\cdot\vec{e}_r\mid c_1,\dots,c_n\in \mathbb F\}=\mathbb F^n$, we have $\forall \vec{q} \in \mathbb V^n: D_y(\vec{q})=1$.
A contradiction with the $1$-Heaviside property.

\end{IEEEproof}

Note that the proof of Lemma \ref{lemma_1_dirac_impossible} has nothing to do with the function $D_x$.
It says that the $1$-Heaviside and $0$-Dirac properties exclude all \emph{next time remedies} for inconsistent decision values $y_i(k)$ in $\mathcal{D}$.
So loosening the requirement for $\mathcal{D}$ to $1$-Dirac property is inevitable, as at least one-round delay is needed in transferring the system state of $\mathcal{D}$.

\subsection{The Dirac constraints}
We say $\mathcal{D}$ is a solution of the $f$-Byzantine broadcast problem upon $K_n$ iff $\mathcal{D}$ satisfies the $1$-Heaviside and $1$-Dirac properties in the presence of $\Upsilon^{[f]}$.
Here, before presenting any concrete solution, we first identify some general design constraints.

Firstly, by the $1$-Dirac property itself, $\mathcal{D}$ should satisfy
\begin{eqnarray}
\label{eq:2Diracs}
\forall \chi\in \Lambda_{D}^{[f]}, k\in \mathbb N:\vec{y}|_{\chi}(k-1)=\vec0 \lor \vec{y}|_{\chi}(k)=\vec1
\end{eqnarray}

Besides, Lemma \ref{lemma_restriction_Dy} polishes (\ref{eq:2Diracs}) by saying that $\vec{y}(k-1)$ can be an arbitrarily valued vector in $\mathbb V^n$ when this $\vec{y}(k-1)\neq \vec0$.
Now, to solve the $f$-Byzantine broadcast problem upon $K_n$, we show the necessity of non-trivial state-transition function $D_x$.
\begin{lemma}
\label{lemma_restriction_Dx}
If $\exists v\in \mathbb V:\forall \vec{q}\in \mathbb V^n: D_x(\vec{q})=v$, no $\mathcal{D}$ can satisfy the $1$-Heaviside and $1$-Dirac properties in the presence of $\Upsilon^{[1]}$.
\end{lemma}
\begin{IEEEproof}
In the case of $v=0$, (\ref{eq:functions_yield_general}) becomes $\vec{y}(k)=D_y(\vec{1}^T\otimes {\vec{u}}(k)+F^{[1]}(k))$.
By applying Lemma \ref{lemma_restriction_Dy}, $\exists \chi \in \Lambda_{D}^{[1]}, \vec{q}\in \mathbb V^n: 0<{\lVert \vec{q} \rVert}<n \land D_y(\hat{\mathbf{X}}|_{\chi}(0))=\vec{q}$.
In round $1$, $\chi$ can take ${\vec{u}}|_{\chi}(1)=\vec0$ and $\mathbf{F}^{[1]}(1)=0$, which results in $\vec{y}|_{\chi}(1)=D_y(0)=0$ by applying Lemma \ref{lemma_restriction_DyDx}.
Since $\vec{y}|_{\chi}(0)\neq 0$, it fails in providing the $1$-Dirac property.

In the case of $v=1$, $\chi$ can take ${\vec{u}}|_{\chi}(k)=\vec0$ and $\mathbf{F}^{[1]}(k)=0$ for all $k\geqslant 0$, with which $\forall k\geqslant 0: \vec{y}|_{\chi}(k)=0$ should hold in providing the $1$-Heaviside property.
But as $D_x(\hat{\mathbf{X}}|_{\chi}(0))=\vec1$, it follows ${\lVert {\vec{x}}|_{\chi}(1)) \rVert}=n$.
With Lemma \ref{lemma_restriction_DyDx}, we have $\vec{y}|_{\chi}(1)=D_y(\hat{\mathbf{X}}|_{\chi}(1))=\vec1$.
A contradiction.

\end{IEEEproof}

Analogous to Lemma \ref{lemma_restriction_Dy}, we can make a similar observation for the state-transition function $D_x$.
\begin{corollary}
\label{co_restriction_Dx}
If $\mathcal{D}$ satisfies the $1$-Heaviside and $1$-Dirac properties, then
$\forall \vec{q}\in \mathbb V^n: \exists \chi \in \Lambda_{D}^{[1]}, k\in \mathbb N: D_x(\hat{\mathbf{X}}|_{\chi}(k))=\vec{q}$.
\end{corollary}
\begin{IEEEproof}
By applying Lemma \ref{lemma_restriction_Dx}, $\exists \vec{q}_0,\vec{q}_1\in \mathbb V^n:D_x(\vec{q}_0)=0 \land D_x(\vec{q}_1)=1$ holds.
Then follow the same steps in proving Lemma \ref{lemma_restriction_Dy}, the conclusion can be drawn similarly.

\end{IEEEproof}

The constraints on $D_y$ can also be refined as follows.
\begin{lemma}
\label{lemma_restriction_Dy2}
If $\mathcal{D}$ satisfies the $1$-Heaviside and $1$-Dirac properties in the presence of $\Upsilon^{[f]}$, then
$\forall \vec{q}\in \mathbb V^n: D_y(\vec{q})=0 \lor {\lVert \vec{q}\rVert}>2f$.
\end{lemma}
\begin{IEEEproof}
Assume there exists a $\vec{p}\in \mathbb V^n$ satisfying ${\lVert \vec{p} \rVert}\leqslant 2f \land D_y(\vec{p})=1$.
As ${\lVert \vec{p} \rVert}\leqslant 2f$, this $\vec{p}$ can always be represented as $\vec{p}=\vec{p}_0+\vec{p}_1$ where $\vec{p}_0,\vec{p}_1\in \mathbb V^n$ and ${\lVert \vec{p}_0 \rVert}+{\lVert \vec{p}_1 \rVert}= {\lVert \vec{p} \rVert} \land {\lVert \vec{p}_0 \rVert}\leqslant f \land {\lVert \vec{p}_1 \rVert}\leqslant f$.
Then, an execution $\chi\in \Lambda_{D}^{[f]}$ can take $D_x(\hat{\mathbf{X}}|_{\chi}(-1))=\vec0$, $\vec{u}|_{\chi}(0)=\vec{p}$ and $\mathbf{F}_{\chi}^{[f]}(0)=(\vec{e}_1-1)^T\otimes \vec{p_0}$ in round $0$.
So we have $\hat{\mathbf{X}}|_{\chi}(0)=\langle \vec{p}, \vec{p}_1, \dots, \vec{p}_1\rangle$.
By applying Lemma \ref{lemma_restriction_DyDx}, we have $D_y(\hat{\mathbf{X}}|_{\chi}(0))=\vec{e}_1$ and ${\lVert D_x(\hat{\mathbf{X}}|_{\chi}(0))\rVert}\leqslant 1$.
Thus $D_y(\hat{\mathbf{X}}|_{\chi}(1))=\vec1$ should hold as is required in (\ref{eq:2Diracs}).
Again with Lemma \ref{lemma_restriction_DyDx}, ${\lVert \hat{\vec{x}}^{(j)}|_{\chi}(1) \rVert}>f$ should holds for all $j\in \{2,\dots,n\}$.
But the same $\chi$ can take $\vec{u}|_{\chi}(1)=\vec{0}$ and $\mathbf{F}_{\chi}^{[f]}(1)=0$ in round $1$, which makes $\vec{x}|_{\chi}(1)=D_x(\hat{\mathbf{X}}|_{\chi}(0))$. So ${\lVert \hat{\vec{x}}^{(j)}|_{\chi}(1) \rVert}={\lVert \vec{x}|_{\chi}(1) \rVert}\leqslant 1$ for all $j\in V$.
A contradiction.

\end{IEEEproof}

And the classical lower-bound of the total number of nodes can be derived \emph{en passant}.
\begin{corollary}
\label{co_restriction_n}
No $\mathcal{D}$ can solve $f$-Byzantine broadcast problem if the total number of local systems is less than $3f+1$.
\end{corollary}
\begin{IEEEproof}
Assume there are only $n\leqslant 3f$ local systems in $\mathcal{D}$.
We choose an execution $\chi\in \Lambda_{D}^{[f]}$ which takes $\vec{u}|_{\chi}(0)=\vec{1}$ in round $0$.
As $\mathcal{D}$ needs to satisfy the $1$-Heaviside property, $\vec{y}|_{\chi}(0)=\vec{1}$ should hold.
But $\chi$ can take $\mathbf{F}_{\chi}^{[f]}(0)=-1^T\otimes \sum_{r=1}^{f}\vec{e_r}$ in round $0$ and makes $\hat{\mathbf{X}}|_{\chi}(0)=1^T\otimes \sum_{r=f+1}^{n}\vec{e_r}$.
Thus ${\lVert \hat{\vec{x}}^{(i)}|_{\chi}(0) \rVert}\leqslant n-f\leqslant 2f$ for all $i\in V$.
Then by applying Lemma \ref{lemma_restriction_Dy2}, $\vec{y}|_{\chi}(0)=\vec{0}$ should be satisfied.
A contradiction.

\end{IEEEproof}

\subsection{A sufficient strategy}
So far, we have identified several constraints (or saying \emph{prerequisites}) on $\mathcal{D}$ in solving the $f$-Byzantine broadcast problem upon $K_n$.
Now we turn to find sufficient $\mathcal{D}$ solutions which can also be referred to as win-strategies in playing the game with the adversary.
These strategies are based on earlier identified constraints and derived with the same system model.

From the perspective of a discrete-time dynamic system, the system-state $\vec{x}$ of $\mathcal{D}$ changes round by round.
These changes can be represented as movements of points (current phases) in the phase-plane.
If we observe these trajectories composed of movements in successive rounds, it would be interesting to find or design some simple patterns.
As the broadcast systems are interfered with by noises, we observe movements of point-sets rather than that of single state-points to describe desired trajectory-patterns.
That is, for any point-set $\Omega_x\subseteq \mathbb V^n$ in the phase-plane, we use $T_x(\Omega_x)$ to represent the point-set of all possible state-points moving from the ones in $\Omega_x$ within a single running step (round) of the dynamic system.
Similarly, we also observe trajectory-patterns of decision signals on the decision-plane by denoting $T_y(\Omega_y)$ as the set of all possible decision vectors that follow the ones in $\Omega_y$ within a single running step too.

Then, the $1$-Dirac property can be expressed as
\begin{eqnarray}
\label{eq:2DiracsEx}
T_y(Y_1)\subseteq Y_2
\end{eqnarray}
by setting $Y_1=\{\vec{y}\in \mathbb V^n\mid \vec{y}\neq \vec0\}$ and $Y_2=\{\vec1\}$.
It says that $\vec{y}(k-1)\in Y_1$ implies $\vec{y}(k)\in T_y(Y_1)\subseteq Y_2$.

Meanwhile, by setting $U_1=\{\vec{u}\in \mathbb V^n\mid \vec{u}\neq \vec0\}$ and $U_2=\{\vec u\in \mathbb V^n\mid{\lVert \vec{u} \rVert}=n\}=\{\vec1\}$, the $1$-Heaviside property can be expressed as the following two implications.
\begin{eqnarray}
\label{eq:1HeavisideEx1}&\vec{y}(k)\in Y_1 &\to \exists k_0\leqslant k: \vec{u}(k_0)\in U_1 \\
\label{eq:1HeavisideEx2}&\vec{u}(k)\in U_2 &\to \vec{y}(k)\in Y_2
\end{eqnarray}

\begin{figure}[htbp]
\centerline{\includegraphics[width=2.0in]{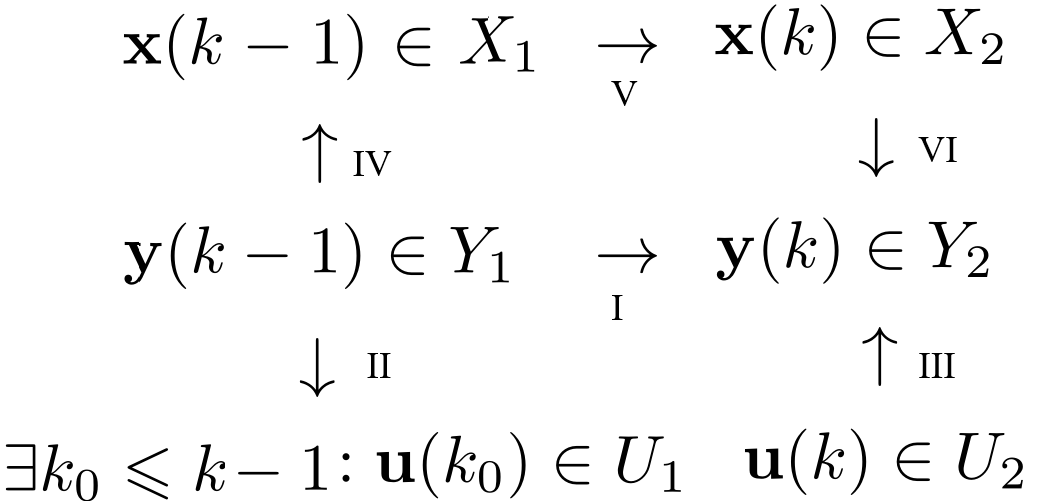}}
\caption{A Sufficient Strategy.}
\label{fig:implication}
\end{figure}

Now to be a solution of the $f$-Byzantine broadcast problem, $\mathcal{D}$ should satisfy (\ref{eq:2DiracsEx}), (\ref{eq:1HeavisideEx1}) and (\ref{eq:1HeavisideEx2}) under $\Upsilon^{[f]}$, which is respectively shown by implications \RNum{1},\RNum{2} and \RNum{3} in Fig.~\ref{fig:implication}.

As is indicated in Lemma \ref{lemma_restriction_Dx}, some non-trivial transition function $D_x$ is indispensable in $\mathcal{D}$ when $f>0$.
In other words, the desired trajectory-pattern on the decision-plane should rely upon the designed trajectory-pattern on the phase-plane.
The trajectory-pattern on the phase-plane can be represented as
\begin{eqnarray}
\label{eq:patternX}
T_x(X_1)\subseteq X_2
\end{eqnarray}
by setting $X_1$ and $X_2$ according to the properties of $D_x$.
By definition, it says that $\vec{x}(k-1)\in X_1$ implies $\vec{x}(k)\in T_x(X_1)\subseteq X_2$, as is shown by implication \RNum{5} in Fig.~\ref{fig:implication}.

Besides, (\ref{eq:patternX}) along is not sufficient to make implication \RNum{1} being true.
Additional implications between some point-sets in both phase-plane and decision-plane should also be drawn.
And these can be represented as
\begin{eqnarray}
\label{eq:phase_decision}
\vec{x}(k)\in X_2 \to \vec{y}(k)\in Y_2\\
\label{eq:decision_phase}
\vec{y}(k)\in Y_1 \to \vec{x}(k)\in X_1
\end{eqnarray}
which is respectively shown by implication \RNum{6} and \RNum{4} in Fig.~\ref{fig:implication}.
As is shown in Fig.~\ref{fig:implication}, implication \RNum{1} would be true if implications \RNum{4},\RNum{5},\RNum{6} are all true.
Thus, implications \RNum{1} to \RNum{3} would be satisfied if implications \RNum{2} to \RNum{6} are all satisfied by some $D_x$, $D_y$, $X_1$ and $X_2$.
For our objective, these implications should be solved as simply and necessarily as possible.

\subsection{The solution}

Now we discuss how to solve $D_x$, $D_y$, $X_1$ and $X_2$ for implications \RNum{2} to \RNum{6}.
Firstly, in the light of identified constraints, we should only consider the \emph{restricted} $\mathcal{D}$ that satisfies all earlier design constraints for being a possible solution.
By applying Lemma \ref{lemma_restriction_DyDx}, \RNum{2} and \RNum{3} hold iff
\begin{eqnarray}
\label{eq:DD0}\forall \vec{q}\in \mathbb V^n: {\lVert \vec{q}\rVert}\leqslant f\to D_y(\vec{q})=D_x(\vec{q})=0 \\
\label{eq:DD1}\forall \vec{q}\in \mathbb V^n: {\lVert \vec1-\vec{q}\rVert}\leqslant f\to D_y(\vec{q})=D_x(\vec{q})=1
\end{eqnarray}
So (\ref{eq:DD0}) and (\ref{eq:DD1}) is necessary and sufficient for implications \RNum{2} and \RNum{3}.

Then, for implications \RNum{4} to \RNum{6}, we expand $\vec{x}$ and $\vec{y}$ as
\begin{eqnarray}
\label{eq:broadcast_x}\vec{x}(k+1)=D_x(\vec{1}^T\otimes\vec{x}(k)+F^{[f]}(k))+{\vec{u}}(k+1) \\
\label{eq:broadcast_y}\vec{y}(k)=D_y(\vec{1}^T\otimes\vec{x}(k)+F^{[f]}(k))
\end{eqnarray}
Now implications \RNum{4}, \RNum{5} and \RNum{6} hold iff
\begin{eqnarray}
\label{eq:implication4}\forall \vec{x}\in \mathbb V^n :D_y(\vec{x}+\vec{\upsilon})=1 \to \vec{x} \in X_1 \\
\label{eq:implication5}\forall \vec{x}\in X_1,\vec{q}\in \mathbb V^n:D_x(\vec{1}^T\otimes \vec{x}+F^{[f]})+\vec{q}\in X_2 \\
\label{eq:implication6}\forall \vec{x}\in X_2:D_y(\vec{x}+\vec{\upsilon}')=1
\end{eqnarray}
where $\vec{\upsilon},\vec{\upsilon}'\in \mathbb F^n$ can take arbitrary values under ${\lVert \vec{\upsilon} \rVert}\leqslant f$ and ${\lVert \vec{\upsilon}' \rVert}\leqslant f$.
Noticing that the column vector $\vec{\upsilon}$ in (\ref{eq:implication4}) is taken from $\mathbf{F}^{[f]}$ in (\ref{eq:implication5}), by combining these two conditions we also have
\begin{eqnarray}
\label{eq:implication45}\forall \vec{x}\in \mathbb V^n,\vec{q}\in \mathbb V^n:\nonumber\\
D_y(\vec{x})=1 \to D_x(\vec{1}^T\otimes \vec{x}+F^{[f]})+\vec{q} \in X_2
\end{eqnarray}
And in deciding a definite $X_1$, it is equivalent to regard the $\vec{\upsilon}$ in (\ref{eq:implication4}) as being absorbed by $\mathbf{F}^{[f]}$ in (\ref{eq:implication5}).
So (\ref{eq:implication4}) simply changes to
\begin{eqnarray}
\label{eq:implication4eqv}\forall \vec{x}\in \mathbb V^n:D_y(\vec{x})=1 \to \vec{x} \in X_1
\end{eqnarray}
By denoting
\begin{eqnarray}
\label{eq:denoteQx} \mathrm Q_x=\{\vec{q}\in \mathbb V^n\mid D_x(\vec{q})=1\} \nonumber\\
\label{eq:denoteQy} \mathrm Q_y=\{\vec{q}\in \mathbb V^n\mid D_y(\vec{q})=1\} \nonumber\\
\label{eq:denoteBf} \mathrm S^n(f)=\{\vec{q}\in \mathbb F^n\mid {\lVert \vec{q} \rVert}\leqslant f\} \nonumber
\end{eqnarray}
where $\mathrm Q_x$ and $\mathrm Q_y$ is respectively the set of nonzero points of $D_x$ and $D_y$ and $\mathrm S^n(f)=\{\vec{\upsilon}\in\mathbb F^n\mid \|\vec{\upsilon}\|\leqslant f\}$ is the $n$ dimensional $0$-norm sphere with radius $f$ in $\mathbb F^n$, (\ref{eq:implication4eqv}) and (\ref{eq:implication6}) can be rewritten as
\begin{eqnarray}
\label{eq:implication4re} \mathrm Q_y \subseteq X_1 \\
\label{eq:implication6re}\forall \vec{x}\in X_2,\vec{\upsilon}\in \mathrm S^n(f):\vec{x}+\vec{\upsilon}\in \mathrm Q_y
\end{eqnarray}
And now Lemma \ref{lemma_restriction_Dy2} and Lemma \ref{lemma_restriction_DyDx} say
\begin{eqnarray}
\label{eq:setQy} \mathrm Q_y \cap \mathrm S^n(2f) = \emptyset \\
\label{eq:setQx} \mathrm Q_x \cap \mathrm S^n(f) = \emptyset \\
\label{eq:setQx2}  \{\vec{q}\in \mathbb V^n\mid{\lVert \vec1-\vec{q} \rVert}\leqslant f\}\subseteq\mathrm Q_x \\
\label{eq:setQy2}  \{\vec{q}\in \mathbb V^n\mid{\lVert \vec1-\vec{q} \rVert}\leqslant f\}\subseteq\mathrm Q_y
\end{eqnarray}
By denoting $\mathrm A \oplus \mathrm B=\{\vec a+\vec b\mid \vec a\in \mathrm A \land \vec b\in \mathrm B\}$, (\ref{eq:implication6re}) to (\ref{eq:setQy2}) are also
\begin{eqnarray}
\label{eq:implication6re2}X_2\oplus \mathrm S^n(f) \subseteq \mathrm Q_y \\
\label{eq:setQyoplus} {\vec0}\notin \mathrm Q_y \oplus \mathrm S^n(2f) \\
\label{eq:setQxoplus} {\vec0}\notin \mathrm Q_x \oplus \mathrm S^n(f) \\
\label{eq:setQx2oplus} \{\vec{1}\}\oplus \mathrm S^n(f)\subseteq\mathrm Q_x \\
\label{eq:setQy2oplus} \{\vec{1}\}\oplus \mathrm S^n(f)\subseteq\mathrm Q_y
\end{eqnarray}
Combine (\ref{eq:implication6re2}) and (\ref{eq:setQyoplus}), we have
\begin{eqnarray}
\label{eq:X2setrange}{\vec0}\notin X_2\oplus \mathrm S^n(f) \oplus \mathrm S^n(2f)=X_2\oplus \mathrm S^n(3f)
\end{eqnarray}
or saying
\begin{eqnarray}
\label{eq:X2setrange2}X_2\cap \mathrm S^n(3f) = \emptyset
\end{eqnarray}

By extending function $D_x$ and the operator $\otimes$ to accept vector-sets as
\begin{eqnarray}
\label{eq:extendingDx}D_x(\mathrm A)=\{D_x(\vec a)\mid \vec a\in \mathrm A\}\\
\label{eq:extendingotimes}\mathrm A\otimes\mathrm B=\{\vec a\otimes\vec b\mid \vec a\in \mathrm A \land \vec b\in \mathrm B\}
\end{eqnarray}
(\ref{eq:implication5}) can be rewritten as
\begin{eqnarray}
\label{eq:implication5re}D_x(\vec{1}^T\otimes X_1 \oplus {\Upsilon}^{[f]})\oplus\mathbb V^n \subseteq X_2
\end{eqnarray}
where (note that here the inconsistent noises can be processed with the extended $D_x$ as the consistent ones)
\begin{eqnarray}
\label{eq:implication5releft}
&&D_x(\vec{1}^T\otimes X_1 \oplus {\Upsilon}^{[f]})\oplus\mathbb V^n \nonumber\\
&=&D_x(\vec{1}^T\otimes X_1 \oplus \vec{1}^T\otimes\mathrm S^n(f))\oplus\mathbb V^n \nonumber\\
&=&D_x(\vec{1}^T\otimes (X_1 \oplus \mathrm S^n(f)))\oplus\mathbb V^n \nonumber\\
&=&\vec{1}\otimes D_x(X_1 \oplus \mathrm S^n(f))\oplus\mathbb V^n
\end{eqnarray}
As (\ref{eq:X2setrange}) says ${\vec0}\notin X_2\oplus \mathrm S^n(3f)$, by (\ref{eq:implication5releft}) we have
\begin{eqnarray}
\label{eq:implication5notin}
{\vec0}\notin \vec{1}\otimes D_x(X_1 \oplus \mathrm S^n(f))\oplus\mathbb V^n
\end{eqnarray}
So we have
\begin{eqnarray}
\label{eq:implication5notinDx}
{\vec0}\notin D_x(X_1 \oplus \mathrm S^n(f))
\end{eqnarray}
As $D_x$ is uniform,
\begin{eqnarray}
\label{eq:implication5inQx}
X_1 \oplus \mathrm S^n(f)\subseteq \mathrm Q_x\\
\label{eq:1inX2}
{\vec1} \subseteq X_2
\end{eqnarray}
should hold for satisfying (\ref{eq:implication5}).
On the contrary, when (\ref{eq:implication5inQx}) and (\ref{eq:1inX2}) hold, (\ref{eq:implication5}) also holds.
So (\ref{eq:implication5inQx}) plus (\ref{eq:1inX2}) is a necessary and sufficient condition for (\ref{eq:implication5}).

Thus, under identified prerequisites, (\ref{eq:implication4}) (\ref{eq:implication5}) (\ref{eq:implication6}) hold iff (\ref{eq:implication4re}) (\ref{eq:implication6re2}) (\ref{eq:implication5inQx}) (\ref{eq:1inX2}) hold.
Combining with identified prerequisites, as (\ref{eq:implication5inQx}) (\ref{eq:implication4re}) (\ref{eq:implication6re2}) (\ref{eq:1inX2}) imply (\ref{eq:setQx2oplus}), (\ref{eq:implication6re2}) (\ref{eq:1inX2}) imply (\ref{eq:setQy2oplus}), and (\ref{eq:setQxoplus}) (\ref{eq:implication5inQx}) (\ref{eq:implication4re}) imply (\ref{eq:setQyoplus}), we can solve $D_x$, $D_y$, $X_1$ and $X_2$ with the following Theorem (whose proof lays above).
\begin{theorem}
\label{theorem_broadcastwholesolution}
$\mathcal{D}$ satisfies implications \RNum{2} to \RNum{6} iff
\begin{eqnarray}
\label{eq:broadcastwholesolution}
{\vec0}\notin \mathrm Q_x \oplus \mathrm S^n(f) \nonumber\\
X_1 \oplus \mathrm S^n(f)\subseteq \mathrm Q_x \nonumber\\
\mathrm Q_y \subseteq X_1 \nonumber\\
X_2\oplus \mathrm S^n(f) \subseteq \mathrm Q_y \nonumber\\
{\vec1} \subseteq X_2
\end{eqnarray}
\end{theorem}

Thus, (\ref{eq:broadcastwholesolution}) is a sufficient solution for the $f$-Byzantine broadcast problem upon $K_n$.
And it also covers the original solution in \cite{SrikanthSimulating} where the corresponding sets can be configured as
\begin{eqnarray}
\label{eq:broadcastsolution0}
\mathrm Q_x&=&\{\vec{q}\in \mathbb V^n\mid{\lVert \vec{q} \rVert}\geqslant n-2f\} \nonumber\\
\mathrm Q_y&=&X_1=\{\vec{q}\in \mathbb V^n\mid{\lVert \vec{q} \rVert}\geqslant n-f\} \nonumber\\
X_2&=&\{\vec{q}\in \mathbb V^n\mid{\lVert \vec{q} \rVert}= n\}
\end{eqnarray}

\section{Broadcast system upon $K_{n_A,n_B}$}
\label{sec:Broadcast_2}
In this section, we consider the broadcast systems $\mathcal{D}_{A,B}$ upon fully connected bipartite network $K_{n_A,n_B}$ and assume an external \emph{General} broadcasts to $V_A$.
The cases where the \emph{General} broadcasts to $V_B$ or belongs to $V$ can be handled similarly.

\subsection{Extended equations}
For fully connected bipartite network, as the adjacency matrix of $K_{n_A,n_B}$ can be represented as $[0,\mathbf{J}_{n_A,n_B};\mathbf{J}_{n_B,n_A},0]$ with $\mathbf{J}_{r,s}$ being the $r\times s$ all-ones matrix, we can specifically represent $\mathcal{D}_{A,B}$ as:
\begin{eqnarray}
\label{eq:functions_transfer2A} &&\vec{x}_A(k)=D_{x_A}(\hat{\mathbf{X}}_B(k-1))+{\vec{u}}(k) \\
\label{eq:functions_estimate2A} &&\hat{\mathbf{X}}_A(k)={\vec 1}_B^T \otimes \vec{x}_A(k)+F_A(k) \\
\label{eq:functions_yield2B}    &&\vec{y}_B(k)=D_{y_B}(\hat{\mathbf{X}}_A(k))\\
\label{eq:functions_transfer2B} &&\vec{x}_B(k)=D_{x_B}(\hat{\mathbf{X}}_A(k)) \\
\label{eq:functions_estimate2B} &&\hat{\mathbf{X}}_B(k)={\vec 1}_A^T \otimes\vec{x}_B(k)+F_B(k) \\
\label{eq:functions_yield2A}    &&\vec{y}_A(k)=D_{y_A}(\hat{\mathbf{X}}_B(k))
\end{eqnarray}
where $\vec x_A$ and $\vec x_B$, $\hat{\mathbf{X}}_A$ and $\hat{\mathbf{X}}_B$, $\vec{y}_A$ and $\vec{y}_A$ are respectively the bipartite state vectors, estimation matrices, decision vectors with respect to $V_A$ and $V_B$.
Obviously, we have $|\vec x_A|=|\vec{y}_A|=|\vec{1}_A|=|V_A|=n_A$ and $|\vec x_B|=|\vec{y}_B|=|\vec{1}_B|=|V_B|=n_B$.
Also, $D_{x_A}$ and $D_{y_A}$ are all uniform in the range of $V_A$.
And $D_{x_B}$ and $D_{y_B}$ are all uniform in the range of $V_B$.
Similar to the set of nonzero points $Q_x$ and $Q_y$ in $\mathcal{D}$, we use $Q_{x_A}$, $Q_{y_A}$, $Q_{x_B}$, $Q_{y_B}$ to respectively denote the corresponding ones of $D_{x_A}$, $D_{y_A}$, $D_{x_B}$, $D_{y_B}$ in $\mathcal{D}_{A,B}$.
And $D_{x_A}$, $D_{y_A}$, $D_{x_B}$, $D_{y_B}$ are also extended to accept vector-sets.
For simplicity, we also use $a$ to represent the set $\{a\}$ when it is not confusing.

Respectively, the dimension of noise matrix $\mathbf{F}_A$ and $\mathbf{F}_B$ is $n_A\times n_B$ and $n_B\times n_A$, which is also the dimension of $\hat{\mathbf{X}}_A$ and $\hat{\mathbf{X}}_B$.
We use $\Upsilon_A^{[f_A]}$ and $\Upsilon_B^{[f_B]}$ to denote the set of all possible $\mathbf{F}_A$ and $\mathbf{F}_B$ in the presence of up to $f_A$ and $f_B$ Byzantine rows.
The column vectors in $\mathbf{F}_A$ and $\mathbf{F}_B$ are still called state noise vectors (or \emph{noises} if it is not confusing).
And the $i$th column vector in $\mathbf{F}_A$ and $\mathbf{F}_B$ is denoted as $\vec{\upsilon}_A^{(i)}$ and $\vec{\upsilon}_B^{(i)}$.
Similar to $\mathrm S^n(f)$ in $\mathcal{D}$, the set of all possible $\vec{\upsilon}_A$ and $\vec{\upsilon}_B$ under $\Upsilon_A^{[f_A]}$ and $\Upsilon_B^{[f_B]}$ is $\mathrm S^{n_A}(f_A)$ and $\mathrm S^{n_B}(f_B)$.
And the set of all possible executions of $\mathcal{D}_{A,B}$ with noises $\mathbf{F}_A\in\Upsilon_A^{[f_A]}$ and $\mathbf{F}_B\in\Upsilon_B^{[f_B]}$ is denoted as $\Lambda_{D_{A,B}}$.

With this, the properties of $D_x$ and $D_y$ may have several equivalent representations.
For example, when $\forall \vec{\upsilon}_B \in \mathrm S^{n_B}(f_B):D_{x_A}(\vec{\upsilon}_B)=0$ holds, the following statements equivalently hold.
\begin{eqnarray}
Q_{x_A}\cap\mathrm S^{n_B}(f_B)=\emptyset \\
D_{x_A}(\mathrm S^{n_B}(f_B))=0 \\
\vec0_B\notin Q_{x_A}\oplus S^{n_B}(f_B)
\end{eqnarray}
And when $\forall \vec{\upsilon}_B \in \mathrm S^{n_B}(f_B):D_{x_A}(\vec1+\vec{\upsilon}_B)=1$ holds, the following statements also equivalently hold.
\begin{eqnarray}
\vec1_B\oplus S^{n_B}(f_B)\subseteq Q_{x_A} \\
D_{x_A}(\vec1_B\oplus\mathrm S^{n_B}(f_B))=1
\end{eqnarray}
And similar equivalent representations also apply to $D_{x_B}$, $D_{y_A}$ and $D_{y_B}$.

\subsection{Heaviside constraints}

Denoting the whole state vector and decision vector in $\mathcal{D}_{A,B}$ as $\vec x=[\vec x_A;\vec x_B]$ and $\vec y=[\vec y_A;\vec y_B]$, the $1$-Heaviside property of $\mathcal{D}_{A,B}$ requires that for every $\chi\in \Lambda_{D_{A,B}}$:
\begin{eqnarray}
\label{eq:Bi1H_1}(\exists k_0\in \mathbb N:\vec{u}\equiv\delta_{+k_0}\cdot \vec1_A) \to \vec{y}\equiv H_{+k_0}\cdot \vec1 \\
\label{eq:Bi1H_0} \vec{u}\equiv 0 \to \vec{y}\equiv 0
\end{eqnarray}

The following constraint is analogous to the ones in $\mathcal{D}$.

\begin{lemma}
\label{lemma_BiDxDy}
If $\mathcal{D}_{A,B}$ satisfies the $1$-Heaviside property, then $D_{y_A}(\vec1)=D_{y_B}(\vec1)=1$, $D_{x_A}(\vec0)=D_{x_B}(\vec0)=D_{y_A}(\vec0)=D_{y_B}(\vec0)=0$ and $D_{x_A}(\hat{\mathbf{X}}_B(-1))= \vec0$.
\end{lemma}
\begin{IEEEproof}

Firstly, for $\chi_1\in \Lambda_{D_{A,B}}$ that makes ${\vec{u}}|_{\chi_1}(0)=\vec{1}_A$ and $\mathbf{F}_{A\chi_1}(0)=0$, $\vec{y}|_{\chi_1}(0)=\vec1$ should hold.
With (\ref{eq:functions_yield2B}), (\ref{eq:functions_estimate2A}) and (\ref{eq:functions_transfer2A}), $\vec{y}_B|_{\chi_1}(0)=D_{y_B}(\hat{\mathbf{X}}_A|_{\chi_1}(0))=D_{y_B}({\vec 1_B}^T \otimes {\vec{x}_A}|_{\chi_1}(0))=D_{y_B}({\vec 1_B}^T \otimes \vec{1}_A)$.
Thus $D_{y_B}(\vec1)=1$ holds.

Secondly, for $\chi_2\in \Lambda_{D_{A,B}}$ which makes ${\vec{u}}|_{\chi_2}(k)=\vec{0}$ and $\mathbf{F}_{A\chi_2}(k)=\mathbf{F}_{B\chi_2}(k)=0$ for all $k\geqslant 0$, $D_{y_B}(D_{x_A}(\hat{\mathbf{X}}_B|_{\chi_2}(k-1)))=0$ should hold for every $k\geqslant 0$.
As $D_{y_B}(\vec1)=1$, we have $D_{x_A}(\hat{\mathbf{X}}_B^{(i)}|_{\chi_2}(k-1))=D_{x_A}({\vec{x}}_B|_{\chi_2}(k-1))\neq 1$ for every $i\in V$ when $k>0$, i.e., $D_{x_A}(\hat{\mathbf{X}}_B^{(i)}|_{\chi_2}(k-1))=D_{x_A}({\vec{x}}_B|_{\chi_2}(k-1))=0$.
Thus $D_{y_B}(\vec0)=0$ holds.
And with (\ref{eq:functions_transfer2A}), we get $\vec{x}_A|_{\chi_2}(k)=\vec0$ for all $k> 0$.
Now as $D_{x_A}({\vec{x}}_B|_{\chi_2}(1))=0$, $D_{x_A}(\vec0)=0$ holds.

Thirdly, if $D_{y_A}(\vec1)=0$, for $\chi_1$, as $\vec{y}|_{\chi_1}(0)=\vec1$ should hold, it follows ${\vec{x}_B}|_{\chi_1}(0)\neq \vec1$, which also means ${\vec{x}_B}|_{\chi_1}(0)= \vec0$.
As $D_{x_A}(\vec0)=0$, it follows ${\vec{x}_A}|_{\chi_1}(1)=\vec0$ and thus $\vec{y}_B|_{\chi_1}(1)=D_{y_B}(\hat{\mathbf{X}}_A|_{\chi_1}(1))=\vec1$ should hold.
It follows $D_{y_B}(\vec0)=1$ which contradicts with $D_{y_B}(\vec0)=0$.

Now as $D_{y_A}(\vec1)=1$, for $\chi_2$, $D_{x_B}(\hat{\mathbf{X}}_A^{(i)}|_{\chi_2}(0))\neq \vec1$ should hold, which means $D_{x_B}(\vec0)=0$ and then $D_{y_A}(\vec0)=0$.

Lastly, as all $D_A^{(i)}$ are uniform in $\mathcal{D}_{A,B}$, $\hat{\mathbf{X}}_B^{(i)}|_{\chi_2}(-1)=\hat{\mathbf{X}}_B^{(j)}|_{\chi_2}(-1)$ should hold for every $i,j\in\{1,2,\dots,n_B\}$.
As $D_{x_A}(\hat{\mathbf{X}}_B|_{\chi_2}(-1))\neq \vec1$, $D_{x_A}(\hat{\mathbf{X}}_B(-1))= \vec0$ holds.

\end{IEEEproof}

Thus, similar to the assumptions in $\mathcal{D}$, we assume that $D_{x_A}(\hat{\mathbf{X}}_B(-1))=\vec0$ holds in the remainder of the paper.

In Lemma \ref{lemma_BiDxDy} and the former Lemmata, constraints are identified by enumerating some special executions of the systems.
Alternatively, with the defined set-operations, we can also directly compute required constraints on the whole.
From now on, we will take this algebraic method.
To compute $1$-Heaviside constraints, now we assume $\vec{u}\equiv\delta_{+k_0}\cdot \vec1_A$ where $k_0\in \mathbb N \cup \{\infty\}$.
By (\ref{eq:Bi1H_1}) and (\ref{eq:Bi1H_0}), $\vec{y}$ should satisfy $\vec{y}(k)= H[k-k_0]\cdot \vec1$ with all $k\geqslant 0$ under $\Upsilon_A^{[f_A]}$ and $\Upsilon_B^{[f_B]}$.
Denoting all possible $\vec{x}(k)$, $\vec{y}(k)$ and $\hat{\mathbf{X}}(k)$ under $\Upsilon_A^{[f_A]}$ and $\Upsilon_B^{[f_B]}$ as $\{\vec{x}(k)\}$, $\{\vec{y}(k)\}$ and $\{\hat{\mathbf{X}}(k)\}$ respectively, it requires
\begin{eqnarray}
\label{eq:BiHy1set} \{\vec{y}(k)\}= H[k-k_0]\cdot \vec1
\end{eqnarray}
Then, by taking (\ref{eq:functions_yield2B}), (\ref{eq:functions_estimate2A}) and (\ref{eq:functions_transfer2A}) into it, we get
\begin{eqnarray}
\label{eq:BiHy1setto}
&&\{\vec{y}_B(k)\} \nonumber\\
&=&\{D_{y_B}(\hat{\mathbf{X}}_A(k))\} \nonumber\\
&=&\{D_{y_B}({\vec 1_B}^T \otimes {\vec{x}_A}(k)+F_A(k))\} \nonumber\\
&=&\{D_{y_B}({\vec 1_B}^T \otimes (D_{x_A}(\hat{\mathbf{X}}_B(k-1))+{\vec{u}}(k))+F_A(k))\} \nonumber\\
&=&D_{y_B}({\vec 1_B}^T \otimes (D_{x_A}(\{\hat{\mathbf{X}}_B(k-1)\})\oplus{\vec{u}}(k)\oplus\mathrm S^{n_A}(f_A)))  \nonumber\\
&=&D_{y_B}(D_{x_A}(\{\hat{\mathbf{X}}_B(k-1)\})\oplus{\vec{u}}(k)\oplus\mathrm S^{n_A}(f_A))\cdot \vec1_B \nonumber\\
&=& H[k-k_0]\cdot \vec1_B
\end{eqnarray}
Thus we have
\begin{eqnarray}
\label{eq:BiHy1settoyB}
D_{y_B}(D_{x_A}(\{\hat{\mathbf{X}}_B(k-1)\})\oplus\delta[k-k_0]\cdot \vec1_A\oplus\mathrm S^{n_A}(f_A))\nonumber\\
 = H[k-k_0]
\end{eqnarray}

For $k=k_0$, (\ref{eq:BiHy1settoyB}) becomes
\begin{eqnarray}
\label{eq:BiHy1settoyBk0}
D_{y_B}(D_{x_A}(\{\hat{\mathbf{X}}_B(k-1)\})\oplus \vec1_A\oplus\mathrm S^{n_A}(f_A)) = 1
\end{eqnarray}
Since $\vec q\oplus \vec1_A = \vec1_A$ holds for all $q\in \mathbb V^{n_A}$, (\ref{eq:BiHy1settoyBk0}) becomes
\begin{eqnarray}
\label{eq:BiHy1settoyBk01}
D_{y_B}(\vec1_A\oplus\mathrm S^{n_A}(f_A)) = 1
\end{eqnarray}

For $k<k_0$, (\ref{eq:BiHy1settoyB}) becomes
\begin{eqnarray}
\label{eq:BiHy1settoyBleqk0}
D_{y_B}(D_{x_A}(\{\hat{\mathbf{X}}_B(k-1)\})\oplus\mathrm S^{n_A}(f_A)) = 0
\end{eqnarray}
Since
\begin{eqnarray}
\label{eq:BiHxsetA}
&&D_{x_A}(\{\hat{\mathbf{X}}_B(k)\}) \nonumber\\
&=&D_{x_A}(\{{\vec 1_A}^T \otimes \vec{x}_B(k)+F_B(k)\}) \nonumber\\
&=&D_{x_A}({\vec 1_A}^T \otimes (\{\vec{x}_B(k)\}\oplus\mathrm S^{n_B}(f_B))) \nonumber\\
&=&D_{x_A}(\{\vec{x}_B(k)\}\oplus\mathrm S^{n_B}(f_B))\cdot \vec1_A
\end{eqnarray}
we have
\begin{eqnarray}
\label{eq:BiHy1settoyBleqk0tox}
D_{y_B}(D_{x_A}(\{\vec{x}_B(k-1)\}\oplus\mathrm S^{n_B}(f_B))\cdot \vec1_A\oplus\mathrm S^{n_A}(f_A)) = 0
\end{eqnarray}
As $D_{y_B},D_{x_A}$ are all uniform, in viewing of (\ref{eq:BiHy1settoyBk01}) we have
\begin{eqnarray}
\label{eq:BiHy1settoyBleqk0tox0}
\{\vec{x}_A(k)\}=D_{x_A}(\{\vec{x}_B(k-1)\}\oplus\mathrm S^{n_B}(f_B)) = 0
\end{eqnarray}
for all $k<k_0$.

Similarly, we have
\begin{eqnarray}
\label{eq:BiHxsetB}
D_{x_B}(\{\hat{\mathbf{X}}_A(k)\})=D_{x_B}(\{\vec{x}_A(k)\}\oplus\mathrm S^{n_A}(f_A))\cdot \vec1_B
\end{eqnarray}
And by taking (\ref{eq:functions_yield2A}), (\ref{eq:functions_estimate2B}) and (\ref{eq:functions_transfer2B}) into (\ref{eq:BiHy1set}), we also have
\begin{eqnarray}
\label{eq:BiHy1settoyA}
D_{y_A}(D_{x_B}(\{\hat{\mathbf{X}}_A(k)\})\oplus\mathrm S^{n_B}(f_B)) = H[k-k_0]
\end{eqnarray}
and thus
\begin{eqnarray}
\label{eq:BiHy1settoyAleqk0tox0}
\{\vec{x}_B(k)\}=D_{x_B}(\{\vec{x}_A(k)\}\oplus\mathrm S^{n_A}(f_A)) = 0
\end{eqnarray}
for all $k<k_0$.

Together with (\ref{eq:BiHy1settoyBleqk0tox0}) and (\ref{eq:BiHy1settoyAleqk0tox0}), we have
\begin{eqnarray}
\label{eq:BiHy1setxA}
D_{x_A}(\mathrm S^{n_B}(f_B)) = D_{x_B}(\mathrm S^{n_A}(f_A)) = 0
\end{eqnarray}
and then by (\ref{eq:BiHy1settoyBleqk0}) and (\ref{eq:BiHy1settoyA}), we have
\begin{eqnarray}
\label{eq:BiHy1setyA}
D_{y_A}(\mathrm S^{n_B}(f_B)) = D_{y_B}(\mathrm S^{n_A}(f_A)) = 0
\end{eqnarray}

Now, for $k\geqslant k_0$, in (\ref{eq:BiHy1set}) it requires
\begin{eqnarray}
\label{eq:BiHy1settoyBgeqk0}
D_{y_B}(D_{x_A}(\{\hat{\mathbf{X}}_B(k-1)\})\oplus\delta[k-k_0]\cdot \vec1_A\oplus\mathrm S^{n_A}(f_A)) = 1
\end{eqnarray}
and
\begin{eqnarray}
\label{eq:BiHy1settoyAgeqk0}
D_{y_A}(D_{x_B}(\{\hat{\mathbf{X}}_A(k)\})\oplus\mathrm S^{n_B}(f_B)) = 1
\end{eqnarray}
Taking (\ref{eq:BiHxsetA}) and (\ref{eq:BiHxsetB}) into (\ref{eq:BiHy1settoyBgeqk0}) and (\ref{eq:BiHy1settoyAgeqk0}), we have
\begin{eqnarray}
\label{eq:BiHy1settoyABgeqk0}
D_{y_B}((D_{x_A}(\{\vec{x}_B(k-1)\}\oplus\mathrm S^{n_B}(f_B))\oplus\delta[k-k_0])\cdot \vec1_A\oplus \nonumber\\
\mathrm S^{n_A}(f_A)) = 1 \\
D_{y_A}(D_{x_B}(\{\vec{x}_A(k)\}\oplus\mathrm S^{n_A}(f_A))\cdot \vec1_B\oplus\mathrm S^{n_B}(f_B)) = 1
\end{eqnarray}
In viewing of (\ref{eq:BiHy1setyA}) we have
\begin{eqnarray}
\label{eq:BiHy1settoyABgeqk011}
0\notin D_{x_A}(\{\vec{x}_B(k-1)\}\oplus\mathrm S^{n_B}(f_B))\oplus\delta[k-k_0] \\
0\notin D_{x_B}(\{\vec{x}_A(k)\}\oplus\mathrm S^{n_A}(f_A))
\end{eqnarray}

As $D_{x_A}$ and $D_{x_B}$ can only take values in $\mathbb V$, we have
\begin{eqnarray}
\label{eq:BiHy1settoyABgeqk0111}
\{\vec{x}_A(k)\}=D_{x_A}(\{\vec{x}_B(k-1)\}\oplus\mathrm S^{n_B}(f_B)) = 1\\
\{\vec{x}_B(k)\}=D_{x_B}(\{\vec{x}_A(k)\}\oplus\mathrm S^{n_A}(f_A)) = 1
\end{eqnarray}
when $k>k_0$.
Thus we get
\begin{eqnarray}
\label{eq:BiHy1settoxyAB1}
D_{y_A}(\vec1_B\oplus\mathrm S^{n_B}(f_B)) = 1 \\
D_{x_A}(\vec1_B\oplus\mathrm S^{n_B}(f_B)) = 1 \\
D_{x_B}(\vec1_A\oplus\mathrm S^{n_A}(f_A)) = 1
\end{eqnarray}

Thus, together with (\ref{eq:BiHy1setxA}), (\ref{eq:BiHy1setyA}), (\ref{eq:BiHy1settoyBk01}), and (\ref{eq:BiHy1settoxyAB1}), we get the following observation.
\begin{lemma}
\label{lemma_Bi1Hf}
$\mathcal{D}_{A,B}$ satisfies the $1$-Heaviside property in the presence of $\Upsilon^{[f]}$ iff
\begin{eqnarray}
D_{x_A}(\mathrm S^{n_B}(f_B)) = D_{x_B}(\mathrm S^{n_A}(f_A)) = 0 \nonumber\\
D_{y_A}(\mathrm S^{n_B}(f_B)) = D_{y_B}(\mathrm S^{n_A}(f_A)) = 0 \nonumber\\
D_{x_A}(\vec1_B\oplus\mathrm S^{n_B}(f_B)) = D_{x_B}(\vec1_A\oplus\mathrm S^{n_A}(f_A)) = 1 \nonumber\\
D_{y_A}(\vec1_B\oplus\mathrm S^{n_B}(f_B)) = D_{y_B}(\vec1_A\oplus\mathrm S^{n_A}(f_A)) = 1 \nonumber
\end{eqnarray}
\end{lemma}
\begin{IEEEproof}
The necessity of the constraints is shown above.
Now we show its sufficiency.
Firstly, we have
\begin{eqnarray}
\{\vec{x}_A(k)\}=D_{x_A}(\{\vec{x}_B(k-1)\}\oplus\nonumber\\
\mathrm S^{n_B}(f_B))\cdot \vec1_A\oplus\delta[k-k_0]\cdot \vec1_A \nonumber\\
\{\vec{x}_B(k)\}=D_{x_B}(\{\vec{x}_A(k)\}\oplus\mathrm S^{n_A}(f_A))\cdot \vec1_B \nonumber
\end{eqnarray}
As $D_{x_A}(\{\vec{x}_B(-1)\}\oplus\mathrm S^{n_B}(f_B))=0$, when $k<k_0$, we have $\{\vec{x}(k)\}=\vec0$.
When $k=k_0$, we have $\{\vec{x}(k)\}=\vec1$.
And when $k>k_0$, we have $\{\vec{x}(k)\}=\vec1$ too.
Thus, we have $\{\vec{x}(k)\}\equiv H[k-k_0]\cdot\vec1$.
Taking this into
\begin{eqnarray}
\{\vec{y}_A(k)\}&=&D_{y_A}(\{\vec{x}_B(k)\}\oplus\mathrm S^{n_B}(f_B))\cdot \vec1_A \nonumber\\
\{\vec{y}_B(k)\}&=&D_{y_B}(\{\vec{x}_A(k)\}\oplus\mathrm S^{n_A}(f_A))\cdot \vec1_B \nonumber
\end{eqnarray}
we get $\{\vec{y}(k)\}\equiv H[k-k_0]\cdot\vec1$.

\end{IEEEproof}

\subsection{A solution}
We say $\mathcal{D}_{A,B}$ is a solution of the $(f_A,f_B)$-Byzantine broadcast problem upon $K_{n_A,n_B}$ iff $\mathcal{D}_{A,B}$ satisfies the $1$-Heaviside and $1$-Dirac properties in the presence of $\Upsilon_A^{[f_A]}$ and $\Upsilon_B^{[f_B]}$.
In the light of the $1$-Heaviside constraints, the $1$-Dirac property should be satisfied under these identified prerequisites.
The $1$-Dirac property requires that for every $\chi\in \Lambda_{D_{A,B}}$:
\begin{eqnarray}
\label{eq:Bidifference}
\exists k'\geqslant 0,\vec{c}\in \mathbb V^n :\Delta \vec{y}(k)\equiv\nonumber\\
\delta[k-k']\cdot \vec{c}+\delta[k-k'-1]\cdot (\vec1-\vec{c})
\end{eqnarray}
where $\Delta \vec{y}(k)=\vec{y}(k)-\vec{y}(k-1)$ is the backward-difference of ${\vec{y}}(k)$.
Analogous to $\mathcal{D}$, to make (\ref{eq:2DiracsEx}) being true, our strategy is to make the implications \RNum{4},\RNum{5},\RNum{6} in Fig.~\ref{fig:implication} being true.
To begin with, we can rewrite (\ref{eq:2DiracsEx}) as follows.
\begin{eqnarray}
\label{eq:Bimplication1}
{\lVert \vec{y}_A(k) \rVert}+{\lVert \vec{y}_B(k) \rVert}> 0 \to \nonumber\\
 (\vec{y}_A(k+1)=\vec1_A \land \vec{y}_B(k+1)=\vec1_B)
\end{eqnarray}

For implication \RNum{6}, with Lemma \ref{lemma_Bi1Hf} we have
\begin{eqnarray}
\label{eq:BiX2setrangeA}{\vec0}\notin X_{A2}\oplus \mathrm S^n_A(f_A) \\
\label{eq:BiX2setrangeB}{\vec0}\notin X_{B2}\oplus \mathrm S^n_B(f_B)
\end{eqnarray}

And implication \RNum{5} can be decomposed into the following four implications.
\begin{eqnarray}
\label{eq:Biimplication5_aa}\vec x_A(k)\in X_{A1} \to \vec x_A(k+1)\in X_{A2} \\
\label{eq:Biimplication5_bb}\vec x_B(k)\in X_{B1} \to \vec x_B(k+1)\in X_{B2} \\
\label{eq:Biimplication5_ba}\vec x_B(k)\in X_{B1} \to \vec x_A(k+1)\in X_{A2} \\
\label{eq:Biimplication5_ab}\vec x_A(k)\in X_{A1} \to \vec x_B(k+1)\in X_{B2}
\end{eqnarray}
With the uniform functions, (\ref{eq:Biimplication5_aa}) to (\ref{eq:Biimplication5_ab}) hold iff
\begin{align}
\label{eq:Biimplication5s_aa}D_{x_A}({\vec 1_A}^T \otimes (D_{x_B}({\vec 1_B}^T \otimes (X_{A1}\oplus\mathrm S^{n_A}(f_A)))\oplus\nonumber\\
\mathrm S^{n_B}(f_B)))\oplus \mathbb V^n \subseteq X_{A2} \\
\label{eq:Biimplication5s_bb}D_{x_B}({\vec 1_B}^T \otimes (D_{x_A}({\vec 1_A}^T \otimes (X_{B1}\oplus\nonumber\\
\mathrm S^{n_B}(f_B)))\oplus\mathbb V^n\oplus\mathrm S^{n_A}(f_A))) \subseteq X_{B2} \\
\label{eq:Biimplication5s_ba}D_{x_A}({\vec 1_A}^T \otimes (X_{B1}\oplus\mathrm S^{n_B}(f_B)))\oplus\mathbb V^n \subseteq X_{A2} \\
\label{eq:Biimplication5s_ab}D_{x_B}({\vec 1_B}^T \otimes (D_{x_A}({\vec 1_A}^T \otimes (D_{x_A}({\vec 1_A}^T \otimes (X_{B1}\oplus\mathrm S^{n_B}\nonumber\\
(f_B)))\oplus\mathrm S^{n_B}(f_B)))\oplus\mathbb V^n\oplus\mathrm S^{n_A}(f_A))) \subseteq X_{B2}
\end{align}

With Lemma \ref{lemma_Bi1Hf} and (\ref{eq:BiX2setrangeA}) and (\ref{eq:BiX2setrangeB}), (\ref{eq:Biimplication5s_aa}) to (\ref{eq:Biimplication5s_ab}) hold iff
\begin{eqnarray}
\label{eq:Biimplication5_XA}X_{A1}\oplus\mathrm S^{n_A}(f_A) \subseteq Q_{x_B} \\
\label{eq:Biimplication5_XB}X_{B1}\oplus\mathrm S^{n_B}(f_B) \subseteq Q_{x_A} \\
\label{eq:Biimplication5_X2A}\vec1 \subseteq X_{A2} \\
\label{eq:Biimplication5_X2B}\vec1 \subseteq X_{B2}
\end{eqnarray}

Like (\ref{eq:implication4re}) and (\ref{eq:implication6re2}) in $\mathcal{D}$, for implication \RNum{4} we have
\begin{eqnarray}
\label{eq:BiQyA} \mathrm Q_{y_A} \subseteq X_{B1} \\
\label{eq:BiQyB} \mathrm Q_{y_B} \subseteq X_{A1}
\end{eqnarray}
And for implication \RNum{6} we have
\begin{eqnarray}
\label{eq:BiQyA2} X_{B2}\oplus\mathrm S^{n_B}(f_B) \subseteq \mathrm Q_{y_A} \\
\label{eq:BiQyB2} X_{A2}\oplus\mathrm S^{n_A}(f_A) \subseteq \mathrm Q_{y_B}
\end{eqnarray}

Thus, with the identified constraints, we can solve $D_{x_A}$, $D_{x_B}$, $D_{y_A}$, $D_{y_B}$, $X_{A1}$, $X_{B1}$, $X_{A2}$ and $X_{B2}$ with the following Theorem (whose proof lays above).
\begin{theorem}
\label{theorem_Bibroadcastwholesolution}
$\mathcal{D}_{A,B}$ satisfies implications \RNum{2} to \RNum{6} iff
\begin{eqnarray}
\label{eq:Bibroadcastwholesolution}
{\vec0}\notin \mathrm Q_{x_A} \oplus \mathrm S^n_B(f_B) \nonumber\\
{\vec0}\notin \mathrm Q_{x_B} \oplus \mathrm S^n_A(f_A) \nonumber\\
X_{A1}\oplus\mathrm S^{n_A}(f_A) \subseteq Q_{x_B} \nonumber\\
X_{B1}\oplus\mathrm S^{n_B}(f_B) \subseteq Q_{x_A} \nonumber\\
\mathrm Q_{y_A} \subseteq X_{B1} \nonumber\\
\mathrm Q_{y_B} \subseteq X_{A1} \nonumber\\
X_{B2}\oplus\mathrm S^{n_B}(f_B) \subseteq \mathrm Q_{y_A} \nonumber\\
X_{A2}\oplus\mathrm S^{n_A}(f_A) \subseteq \mathrm Q_{y_B} \nonumber\\
\vec1 \subseteq X_{A2} \nonumber\\
\vec1 \subseteq X_{B2}
\end{eqnarray}
\end{theorem}

Thus, (\ref{eq:Bibroadcastwholesolution}) is a sufficient solution for the $(f_A,f_B)$-Byzantine broadcast problem upon $K_{n_A,n_B}$.
And analogous to (\ref{eq:broadcastsolution0}), a simple concrete solution can be configured as
\begin{eqnarray}
\label{eq:Bibroadcastsolution0}
\mathrm Q_{x_A}&=&\{\vec{q}\in \mathbb V^{n_B}\mid{\lVert \vec{q} \rVert}\geqslant n_B-2f_B\} \nonumber\\
\mathrm Q_{x_B}&=&\{\vec{q}\in \mathbb V^{n_A}\mid{\lVert \vec{q} \rVert}\geqslant n_A-2f_A\} \nonumber\\
\mathrm Q_{y_A}&=&X_{B1}=\{\vec{q}\in \mathbb V^{n_B}\mid{\lVert \vec{q} \rVert}\geqslant n_B-f_B\} \nonumber\\
\mathrm Q_{y_B}&=&X_{A1}=\{\vec{q}\in \mathbb V^{n_A}\mid{\lVert \vec{q} \rVert}\geqslant n_A-f_A\} \nonumber\\
X_{A2}&=&\{\vec{q}\in \mathbb V^{n_A}\mid{\lVert \vec{q} \rVert}= n_A\}\nonumber\\
X_{A2}&=&\{\vec{q}\in \mathbb V^{n_B}\mid{\lVert \vec{q} \rVert}= n_B\}
\end{eqnarray}

In Fig.~\ref{fig:algo_broadcast}, we present an algorithm $\mathtt{BI\_BROADCAST}$ corresponding to (\ref{eq:Bibroadcastsolution0}).
In this algorithm, to support parallel broadcasts in the agreement systems (in Section~\ref{sec:Agreement_Ex}), the broadcasted message could contain a multi-valued \emph{General} identifier $g$ (later in Section~\ref{sec:app} we would also discuss how to substitute these identifiers with TDMA communication).
And the state variable $x_i$ and the estimated system state vector $\hat{x}^{(i)}$ in each node $i\in V$ are also extended as \emph{General}-indexed arrays.
We use $send(i,j,g,v,k)$ and $recv(i,j,g,v,k)$ to respectively represent basic communication primitives for sending and receiving a message $(g,v)$ from node $i$ to node $j\in N_i$ at round $k$, where $g$ is the specific \emph{General} and $v\in \mathbb V$ is the Boolean value.
Following the general assumption, here we also assume a message with a \emph{General} identifier can be sent, received, and processed in the same round.
The broadcast system is executed by operating three functions.
The function $init(i,g,v)$ initiates node $i$ for the \emph{General} $g$ with the initial value $v$ (for $i\in V_A$ only).
The function $bcast(i,g,v,k)$ distributes a message $(g,v)$ from node $i$ to $N_i$ at round $k$.
The function $accept(i,g,v,k)$ accepts (with the decision value $y_i(k)=1$) the message $(g,v)$ in node $i$ at round $k$.
For efficiency, we also use a variable $b[g]$ to record the current local state for the \emph{General} $g$.
With $b[g]$ (and the state variable and the estimated system state vector), as each correct node is excited (i.e., when $x_i$ becomes $1$) at most once (as the state signal is monotonically increasing) for a \emph{General} during an execution, the required overall traffic in an execution of $\mathtt{BI\_BROADCAST}$ can be bounded by $2n_A n_B \lceil\log_2 m\rceil$ bits if there is at most $m$ \emph{Generals}.

\alglanguage{pseudocode}
\algrenewcommand{\algorithmiccomment}[1]{\hskip1em//#1}
\begin{figure}[htbp]
\centering
\subfloat[For Node $i\in V_A$\label{algo:broadcast-A}]{
\begin{minipage}{.22\textwidth}
\centering
\begin{algorithmic}[1]
\Statex {\underline{$init(i,g,v)$}:}
    \State  $b[g]:=x_i[g]:=v$;
    \State  $\hat{\vec x}_B^{(i)}[g]:=\vec0$;
    \State  $bcast(i,g,x_i[g],0)$
\Statex
\Statex {\underline{$bcast(i,g,v,k)$}:}
    \ForAll {$j\in V_B$}
    \State  $send(i,j,g,v,k)$;
    \EndFor
\Statex
\Statex {\underline{on $recv(j,i,g,v,k)$}:}
    \State  $\hat{x}_j^{(i)}[g]:=v$;
    \If{${\lVert \hat{\vec x}_B^{(i)}[g] \rVert} \geqslant n_B-2f_B$}
        ~  $x_i[g]:=1$;
    \EndIf
    \If{$x_i[g]=1 \land \lnot b[g]$}
        \State  $b[g]:=1$;
        \State  $bcast(i,g,1,k+1)$;
    \EndIf
    \If{${\lVert \hat{\vec x}_B^{(i)}[g] \rVert} \geqslant n_B-f_B$}
        ~  $accept(i,g,v,k)$;
    \EndIf
\end{algorithmic}
\end{minipage}}
\hfill
\subfloat[For Node $j\in V_B$\label{algo:broadcast-B}]{
\begin{minipage}{.22\textwidth}
\centering
\begin{algorithmic}[1]
\Statex {\underline{$init(j,g,0)$}:}
    \State  $b[g]:=x_j[g]:=0$;
    \State  $\hat{\vec x}_A^{(j)}[g]:=\vec0$;
    \Statex
\Statex
\Statex {\underline{$bcast(j,g,v,k)$}:}
    \ForAll {$i\in V_A$}
    \State  $send(j,i,g,v,k)$;
    \EndFor
\Statex
\Statex {\underline{on $recv(i,j,g,v,k)$}:}
    \State  $\hat{x}_i^{(j)}[g]:=v$;
    \If{${\lVert \hat{\vec x}_A^{(j)}[g] \rVert} \geqslant n_A-2f_A$}
        ~  $x_j[g]:=1$;
    \EndIf
    \If{$x_j[g]=1 \land \lnot b[g]$}
        \State  $b[g]:=1$;
        \State  $bcast(j,g,1,k)$;
    \EndIf
    \If{${\lVert \hat{\vec x}_A^{(j)}[g] \rVert} \geqslant n_A-f_A$}
        ~  $accept(j,g,v,k)$;
    \EndIf
\end{algorithmic}
\end{minipage}}
\caption{The $\mathtt{BI\_BROADCAST}$ Algorithm.}\label{fig:algo_broadcast}
\end{figure}

\section{Some extensions upon bipartite bounded-degree networks}
\label{sec:Broadcast_3}
In the former section, we have investigated the broadcast problem upon fully connected bipartite networks.
In this section, we investigate this upon bipartite bounded-degree networks.

\subsection{A simulation upon bipartite networks}
Here we first give a general bipartite solution by explicitly constructing some bipartite networks to simulate the existing BFT solution provided upon butterfly networks in \cite{Dwork1986}.

The simulation of the protocols originally designed for butterfly networks upon the corresponding bipartite networks is straightforward.
That is, for an $r$-butterfly network $G_{bu}=(V_{bu},E_{bu})$ (see \cite{Dwork1986}) with $r=\log_2 s$ distinct layers (denoted as $L_i$ for $0\leqslant i< r$) and $s$ distinct nodes (denoted as $v_{i,j}$ for $0\leqslant j< s$) in each $L_i$, we can construct a corresponding bipartite network $G_{bi}=(V_0\cup V_1,E)$ with $|V_0|=|V_1|=s$ as follow (for simplicity here we assume $r$ is an even integer).
Firstly, in $G_{bi}$, for each $i\in\{0,1\}$, denote the $j$th node in $V_i$ as $v_i(j)$ with $0\leqslant j<s$.
Then, set $(v_0(j_0),v_1(j_1))\in E$ iff $\exists i,i_0,i_1\in\mathbb Z:i_0= (2i\bmod r) \land i_1=((2i+1)\bmod r) \land (v_{i_0,j_0},v_{i_1,j_1})\in E_{bu}$.
Similarly, set $(v_1(j_1),v_0(j_0))\in E$ iff $\exists i,i_0,i_1\in\mathbb Z:i_1=((2i-1)\bmod r) \land i_0=(2i\bmod r) \land (v_{i_1,j_1},v_{i_0,j_0})\in E_{bu}$.
Then, each node $v_{i,j}$ with an even $i$ in the butterfly network $G_{bu}$ is simulated by the node $v_0(j)$ in the homeomorphic bipartite network $G_{bi}$.
And each node $v_{i,j}$ with an odd $i$ is simulated by $v_1(j)$.
Thus, each node $v_b(j)$ of $G_{bi}$ (with $b\in \{0,1\}$) simulates exactly $r/2$ nodes $v_{i,j}$ (with $i\equiv b\bmod 2$) of $G_{bu}$.

Denoting the original fault-tolerant system built upon the butterfly networks as $\mathcal{Y}$ and the corresponding simulation system upon the bipartite networks as $\mathcal{B}$, we show that $\mathcal{B}$ can be built upon logarithmic-degree bipartite networks.

\begin{theorem}
\label{theorem_broadcast_polylogarithmic}
An $f$-Byzantine $\epsilon$-incomplete system $\mathcal{Y}$ upon an $r$-butterfly network ($r$ is even) can be simulated in an $(\Omega(f/\log n),\Omega(f/\log n))$-Byzantine $O(\epsilon\log n)$-incomplete system $\mathcal{B}$ upon some logarithmic-degree bipartite network.
\end{theorem}
\begin{IEEEproof}
Assume $\mathcal{Y}$ is built upon the $r$-butterfly network $G_{bu}$ and thus we can construct the corresponding bipartite network $G_{bi}$.
For our aim, we assume there are up to $f/r$ Byzantine nodes in each side of $G_{bi}$.
As $r$ is even, the simulated $\mathcal{Y}$ system can be viewed as a $\mathcal{Y}$ upon $G_{bu}$ in the presence of up to $f$ Byzantine nodes.
With the Theorem~3 of \cite{Dwork1986}, an $f$-Byzantine resilient $O(f(\log f-1))$-incomplete system $\mathcal{Y}$ exists upon $G_{bu}$.
And as $G_{bu}$ is $4$-regular, the bipartite network $G_{bi}$ is $(d=2r)$-regular.
As there are $n=2s=2^{r+1}$ nodes in $G_{bi}$, we have $d=O(\log n)$.
So there is an $(\Omega(f/\log n),\Omega(f/\log n))$-Byzantine $O(f(\log f-1)\log n)$-incomplete system $\mathcal{B}$ upon the $O(\log n)$-degree $G_{bi}$.
\end{IEEEproof}

So it is clear that the Byzantine-tolerant incomplete systems can be built upon logarithmic-degree bipartite networks.
Actually, the good properties of the butterfly networks can also be largely gained by simulating the butterfly networks in non-bipartite logarithmic-degree networks (just to further combine the two layers of $G_{bi}$).
Here we argue that as the two sides of the bipartite networks can often be interpreted as the commutating components and the communicating components in real-world systems, each communication round in such systems can be divided into two fault-tolerant processing phases in the bipartite networks, with which extra efficiency is gained.

It should be noted that, however, the resilience $\alpha$ (with which $f$ should be no more than $\alpha n$ in an $n$-nodes system) is not a constant number in the $\mathcal{Y}$ systems.
Denoting the resilience of $\mathcal{Y}$ as $\alpha_{bu}$, \cite{Dwork1986} shows that the $\mathcal{Y}$ system can only achieve $\alpha_{bu} =\Omega(1/\log n)$.
We can see that the resilience of $\mathcal{B}$ performs no worse than that of $\mathcal{Y}$, but also no better.

\subsection{A general extension for solutions upon bipartite expanders}
For bipartite expanders, as the result developed in \cite{ALON198815,Upfal1992,UPFAL1994312} only work in the non-bipartite settings, the natural extension of these works are not straightforward.
Nevertheless, denoting the adjacency matrix of a $(d_0,d_1)$-biregular \citep{Spielman2015I,Spielman2015IV} bipartite expander $G=(V_0\cup V_1,E)$ as $A$, by performing a two-phase communication round upon $G$, we get a disconnected graph $G'$ with the adjacency matrix $A^2=[A_0,0;0,A_1]$.
As the original bipartite expander \citep{RN4359} is connected, $A_0$ and $A_1$ are the adjacency matrices of two connected subgraphs of $G'$, denoted as $G_0$ and $G_1$.
And each adjacency matrix $A_i$ with $i\in \{0,1\}$ now has a unique eigenvalue with the maximal absolute value $d=d_0 d_1$ that corresponds to the eigenvector $\vec 1$.
By applying the spectral theorem \citep{Bapat2014}, we can easily see that all the other eigenvalues of $A_0$ (and $A_1$) are non-negative and no more than $O(d_0+d_1)$.
And as $G_0$ and $G_1$ are $d$-regular, with Lemma~2.3 of \cite{ALON198815} we still have
\begin{eqnarray}
\label{eq:ramanujan2}
|e(S_i) -\theta^2dn_i/2|\leqslant O(d^{1/2})\theta(1-\theta)n_i/2~~
\end{eqnarray}
with $i\in \{0,1\}$, $S_i\subseteq V_i$, $|S_i|=\theta$ and $e(S_i)$ being the number of the internal edges of the subgraph of $G_i$ induced by $S_i$.
Notice that the number of the edges in $G_0$ (and $G_1$) are also increased to $n_0 d/2$ (and $n_1 d/2$) with $d=d_0 d_1$, there still exists non-trivial restraint condition on the number of the edges of $S_0$ (or $S_1$).
Concretely, for every $T_0\subseteq V_0$ and $T_1\subseteq V_1$ with $|T_0|\leqslant f_0$ and $|T_1|\leqslant f_1$, denoting the set of all npc nodes in $G_i$ as $P_i(T_0,T_1,\beta_0,\beta_1)$ for $i\in \{0,1\}$, $P_i(T_0,T_1,\beta_0,\beta_1)$ can be constructed in the $\mathtt{NPC}$ procedure just in the similar way of \cite{Upfal1992}.

\alglanguage{pseudocode}
\algrenewcommand{\algorithmiccomment}[1]{\hskip1em//#1}
\begin{figure}[htbp]
\centering
\begin{algorithmic}[1]
\Procedure{$\mathtt{NPC}$}{$G,\beta_0,\beta_1,T_0,T_1$}:
    \State $Z_0:=Z_1:=\emptyset$;
    \State $Z_0':=\{j\in V_0\mid |N_j\cap (T_1\cup Z_1)|\geqslant \beta_0 d_0 \}$;
    \State $Z_1':=\{j\in V_1\mid |N_j\cap (T_0\cup Z_0)|\geqslant \beta_1 d_1 \}$;
    \While {$Z_0'\cup Z_1'\neq \emptyset$}
        \State  $Z_0:=Z_0\cup Z_0'$;
        ~  $Z_1:=Z_1\cup Z_1'$;
        \State $Z_0':=\{j\in V_0\mid |N_j\cap (T_1\cup Z_1)|\geqslant \beta_0 d_0 \}$;
        \State $Z_1':=\{j\in V_1\mid |N_j\cap (T_0\cup Z_0)|\geqslant \beta_1 d_1 \}$;
    \EndWhile
    \State $P_0:=V_0\setminus (Z_0\cup T_0)$;
    ~ $P_1:=V_1\setminus (Z_1\cup T_1)$;
    \State  \Return $(P_0(T_0,T_1,\beta_0,\beta_1),P_1(T_0,T_1,\beta_0,\beta_1))$;
\EndProcedure
\end{algorithmic}
\caption{The $\mathtt{NPC}$ Procedure.}\label{fig:algo_npc2}
\end{figure}

Now we show that the basic result of \cite{ASBBDNFTP} can be extended to bipartite expanders, i.e., the sets $P_i(T_0,T_1,\beta_0,\beta_1)$ would have sufficient sizes for specific $\beta_0$ and $\beta_1$.

\begin{lemma}
\label{lemma_sufficient_largeX}
Denoting the second large absolute value of the eigenvalues of $A_i$ as $\lambda_i$, for any $\alpha_i,\beta_i\in (0,1)$, if
\begin{eqnarray}
\label{eq:necessary_largeX}
\beta_i-\sqrt{2\alpha_i\beta_i}\geqslant \lambda_i/(2d)
\end{eqnarray}
then there exists $\mu< \sqrt{2\beta_i/\alpha_i}$ satisfying $\forall T\subset V_i: |T|\leqslant\alpha_i |V_i| \to |P_i(T_0,T_1,\beta_0,\beta_1)|> |V_i|-\mu |T_i|$.
\end{lemma}
\begin{IEEEproof}
For any $i\in\{0,1\}$, let $P_i=P_i(T_0,T_1,\beta_0,\beta_1)$ and $|P_i|=|V_i|-\mu_i|T_i|$.
For every $\mu\in(1,\mu_i)$, as the subgraph of $G_i$ induced by any subset $S_i\subseteq V_i\setminus P_i$ with $|S_i|=\mu |T_i|$ has at least $(\mu-1)|T_i|\beta_i d_i$ internal edges, with Lemma~2.3 of \cite{ALON198815}, $|\beta_i(\mu-1)/\mu-\alpha_i\mu /2|<\lambda_i/(2d)$ holds.
Now denote $g(x)=\beta_i(x-1)/x-\alpha_i x /2$ and suppose $\mu_i\geqslant\sqrt{2\beta_i/\alpha_i}$.
As $g(\sqrt{2\beta_i/\alpha_i})=\beta_i-\sqrt{2\alpha_i\beta_i}$, $\beta_i-\sqrt{2\alpha_i\beta_i}<\lambda_i/(2d)$ holds.
A contradiction with (\ref{eq:necessary_largeX}).
\end{IEEEproof}

Now as $\lambda_i=O(d_0+d_1)$ with $d=d_0 d_1$, $\lambda_i/(2d)\to 0$ with $d_0\to \infty$ and $d_1\to \infty$.
In other words, the basic asymptotical relation between $e(S_i)$, $|S_i|$, $d$, $\beta_i$ and $\mu_i$ remains unchanged.
Meanwhile, the relation between the new $d$ and $n_i$ are changed in the favor of the node-degrees for the original bipartite expander $G$.
So the solutions upon non-bipartite expanders can be generally transformed and applied (might even better) to any side of the bipartite expanders with only minor technical differences.
In considering that the two sides of the bipartite networks can often be interpreted as computing components and communicating components in real-world systems, the solutions provided upon the bipartite bounded-degree networks may have special usefulness.

\subsection{Finer properties of biregular bounded-degree networks}
The extension of solutions upon non-bipartite networks for the ones upon bipartite networks discussed above is a simple strategy.
But it only makes use of the connectivity properties of just one side of a biregular network.
A finer property of biregular networks can also be developed by directly extending the basic result of \cite{ALON198815}.
Now assume $G=(V_0\cup V_1,E)$ being a $(d_0,d_1)$-biregular bipartite graph with two sides $V_0$ and $V_1$ (also denote $n_0=|V_0|$, $n_1=|V_1|$ and $n=|V|$).
Denote $A$ as the adjacency matrix of $G$.
And by excluding one largest eigenvalue and one smallest eigenvalue of the $A$ (the eigenvalues may be multiple), denote $\lambda$ as the largest absolute value of the remaining $n-2$ eigenvalues.
Then the result of Lemma~2.3 of \cite{ALON198815} can be extended for biregular networks (and also multi-regular ones if needed).

\begin{lemma}[extending \citep{ALON198815}]
\label{lemma_ramanujan_bipartite}
For every $S_0\subseteq V_0$ and $S_1\subseteq V_1$ with $|S_0|=\theta_0 n_0>0$ and $|S_1|=\theta_1 n_1>0$,
\begin{eqnarray}
\label{eq:ramanujan_bipartite}
|e(S_0,S_1) -\frac{1}{2}\theta_0\theta_1 (d_0 n_0+d_1 n_1)|\leqslant~~~~~~~~~~\nonumber\\
\frac{\lambda}{2}[\theta_1(1-\theta_0)n_0+\theta_0(1-\theta_1)n_1]
\end{eqnarray}
holds, where $e(S_0,S_1)=|E\cap (S_0\times S_1)|$ is the number of the internal edges of the subgraph of $G$ induced by $S_0\cup S_1$.
\end{lemma}
\begin{IEEEproof}
Analogous to the proof of Lemma~2.3 of \cite{ALON198815}, now define the vector $f:V\to \mathbb R$ by $f(i)=(n_0-|S_0|)/|S_0|$ if $i\in S_0$, $f(i)=(n_1-|S_1|)/|S_1|$ if $i\in S_1$, and $f(i)=-1$ if $i\in V\setminus (S_0\cup S_1)$.
Now since $\sum_{i=1}^n f(i)=0$ and $\sum_{i=1}^n f(i)(b_{i,V_0}-b_{i,V_1})=0$ with $b_{a,A}=1$ if $a\in A$ and $b_{a,A}=0$ otherwise, i.e., $f$ is orthogonal to the eigenvectors of the largest eigenvalue and the smallest eigenvalue of the adjacency matrix of $G$, $|(Af,f)|\leqslant \lambda (f,f)$ still holds, i.e., $|\sum_{i,j\in E}(f(i)-f(j))^2-d\sum_{i=1}^n f^2(i)|\leqslant \lambda \sum_{i=1}^n f^2(i)$.

As now we have $\sum_{i,j\in E}(f(i)-f(j))^2=e(S_0,S_1)(1/\theta_0-1/\theta_1)^2+e(S_0,V_1\setminus S_1)(1/\theta_0)^2+e(V_0\setminus S_0, S_1)(1/\theta_1)^2$ and $e(S_0,V_1\setminus S_1)=d_0|S_0|-e(S_0,S_1)$ and $e(V_0\setminus S_0,S_1)=d_1|S_1|-e(S_0,S_1)$, we get $\sum_{i,j\in E}(f(i)-f(j))^2=d_0 n_0/\theta_0+d_1 n_1/\theta_1-2e(S_0,S_1)/(\theta_0 \theta_1)$.
So, with $\sum_{i=1}^n f^2(i)=n_0/\theta_0+n_1/\theta_1-n$, we have $|d_0 n_0/\theta_0+d_1 n_1/\theta_1-2e(S_0,S_1)/(\theta_0 \theta_1)-d_0 n_0/\theta_0-d_1 n_1/\theta_1+d_0 n_0+d_1 n_1|\leqslant \lambda (n_0/\theta_0+n_1/\theta_1-n)$ and thus the conclusion holds.
\end{IEEEproof}

Notice that (\ref{eq:ramanujan_bipartite}) is a property of any subset $S\subseteq V$ that contains the nodes in both sides of $G$.
Now view the right side of (\ref{eq:ramanujan_bipartite}) as the uncertainty of the number of edges between $S_0$ and $S_1$ and denote it as $\Delta e(\theta_0,\theta_1)$.
When $\theta_0=\theta_1$, $S$ has the same property as that in non-bipartite networks, where $\Delta e(\theta_0,\theta_1)$ reaches its peak at $\theta_0=\theta_1=1/2$.
When $\theta_0\neq\theta_1$, $\Delta e(\theta_0,\theta_1)$ increases with the increase of $\theta_0$ if $\theta_1<n_1/n$.
But if $\theta_1>n_1/n$, then $\Delta e(\theta_0,\theta_1)$ decreases with the increase of $\theta_0$.
This means that when $|S_1|$ is sufficiently large (larger than $n_1^2/n$), the uncertainty of the number of edges between $S_0$ and $S_1$ would decrease with the increase of $|S_2|$.
And this is a finer property than that in non-bipartite networks.

With this, we see that the basic strategies taken in the last section can also be applied in general strong enough biregular bipartite expanders.

\section{The agreement systems under a general framework}
\label{sec:Agreement_Ex}
\begin{figure}[htbp]
\centering
\subfloat[Abstraction\label{fig:agreement_system1}]{\centering\includegraphics[width=0.9in]{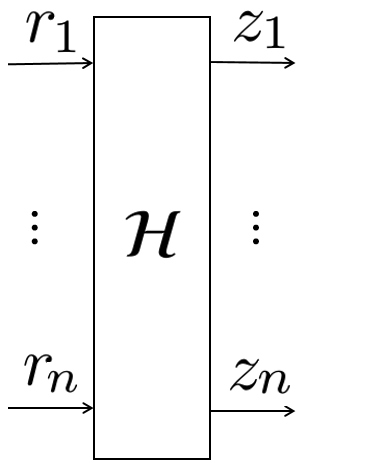}}
\subfloat[The Internal Structure\label{fig:agreement_system2}]{\centering\includegraphics[width=2.4in]{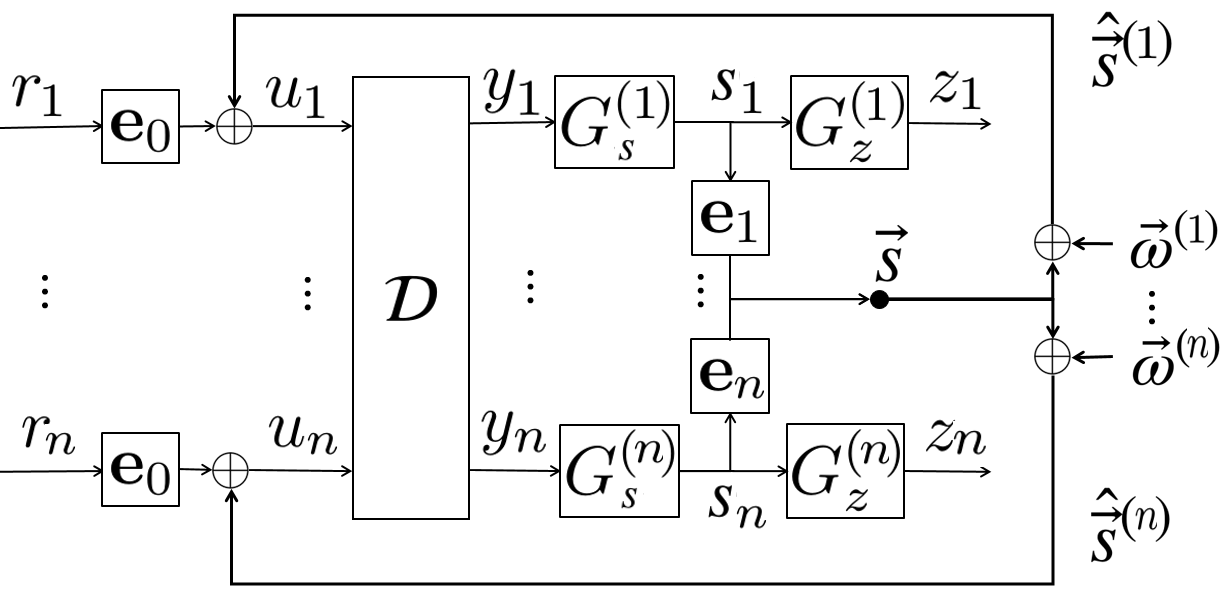}}
\caption{A Modularized Agreement System.}
\end{figure}

In this section, we provide a framework in converting the general relay-based broadcast systems into the corresponding agreement systems.
As is shown in Fig.~\ref{fig:agreement_system1}, a Byzantine agreement system (\emph{agreement system} for short) $\mathcal{H}$ transfers the reference signal $\vec{r}$ (from the top-rank \emph{General}) to the final agreed signal $\vec{z}$ (for this \emph{General}).
Different from the relay-based broadcast system, the execution of an agreement system should always terminate within bounded rounds.

Generally, the BA solutions can be immediate or eventual \citep{Dolev1982}.
For simplicity, here we focus on immediate agreements.
We can see that the corresponding eventual ones can also be derived in the same framework easily.
Now, the classical immediate agreement requires that
\begin{eqnarray}
\label{eq:AgreeH_a}
\textit{Agreement}&:&\forall k:\exists v\in \mathbb V:\vec{z}(k)= v\cdot\vec1 \nonumber\\
\label{eq:AgreeH_v}
\textit{Validity}&:&\forall v\in \mathbb V: \vec{r}\equiv v\delta\cdot \vec1 \to\vec{z}(k_f)\equiv v \cdot \vec1 \nonumber
\end{eqnarray}
where $k_f\in \mathbb N$ is a fixed round number.
Thus, all correct nodes can terminate at round $k_f$, which is also referred to as the \emph{simultaneity} property in \cite{RN4346}.
Here as we also consider incomplete solutions, we allow the agreement to be reached in $(1-\mu\alpha)n$ npc nodes.

\subsection{Modularized agreement systems}
With the provided broadcast systems, the agreement system can have a modularized structure.
Based on the general broadcast system $\mathcal{D}$ upon general networks, the structure of $\mathcal{H}$ is shown in Fig.~\ref{fig:agreement_system2}.
According to \cite{Toueg1987Fast}, the $\mathcal{D}$-based agreement system can be represented as:
\begin{eqnarray}
\label{eq:agree_functions_transfer}&&\vec{y}^{(i)}(k)=D(\hat{\vec{s}}^{(i)}(k-1)+r_i(k)\vec{e}_0) \\
\label{eq:agree_functions_estimate}&&\hat{\vec{s}}^{(i)}(k)=\vec{s}(k)+\vec{\omega}^{(i)}(k) \\
\label{eq:agree_functions_yield}&&s_i(k)=G_s^{(i)}(\vec{y}^{(i)}(k),k) \\
\label{eq:agree_functions_agree}&&z_i(k)=G_z^{(i)}(s_i(k),k)
\end{eqnarray}
where $k\geqslant 0$ and $\vec{s}(k)=s_1(k)\vec{e}_1+s_2(k)\vec{e}_2+\dots+s_n(k)\vec{e}_n$ is the \emph{current decision} vector which is comprised of the current decision values in BA algorithms of \cite{Toueg1987Fast} except for that we not employ the early-stopping operations for simplicity.
In this simplified system, each $i\in V$ has a current decision value $s_i(k)$ to indicate the current decision of the agreement in round $k$.
And these current decisions would be held in block $G_z^{(i)}$ and not be output as the \emph{final decisions} $z_i(k)$ until the specific round $k=k_f$, at which the executions in all correct (or at least npc in the case of incomplete solutions, the same as below) nodes can terminate simultaneously.
Eventual agreements can also be built without $G_z^{(i)}$ for reaching an agreement sometimes earlier in $\mathcal{H}$, but which is not much helpful in hard-real-time applications.

Now, to reach an immediate agreement at the $k_f$th round in the correct nodes, each decision value $s_i$ also acts as feedback to indicate other nodes about the current decision in each node $i$.
Since there can be Byzantine decision indicators, we use a decision noise vector $\vec{\omega}^{(i)}(k)={\omega}_1^{(i)}(k)\vec{e}_1+\dots+{\omega}_n^{(i)}(k)\vec{e}_n$ to equivalently represent the effect of faulty decisions indicated to the node $i$ in round $k$, where $\omega_j^{(i)}(k)\in \mathbb F$ can be inconsistent just like the noises in $\mathcal{D}$.
Then, these interfered decision signals are fed back to the input signals of the broadcast system $\mathcal{D}$ (i.e., the $u$ signals), together with the system input signal $\vec r$.
Then, these noisy $u$ signals are independently filtered in the broadcast system $\mathcal{D}$ and then to be respectively yielded as independent $y$ signals in $\mathcal{D}$.
That is, the input signal $\vec{u}_i$ of each local system $D^{(i)}$ is no longer a Boolean value in each round $k$.
Instead, each input signal $u_i$ is comprised of $n+1$ Boolean values from the $n$ distinct nodes and the assumed external \emph{General}.
As these $n+1$ values are orthogonal, we can use an $n+1$ dimensional vector $r_i(k)\vec{e}_0+\hat s_{1,i}(k-1)\vec{e}_1+\hat s_{2,i}(k-1)\vec{e}_2+\dots+\hat s_{n,i}(k-1)\vec{e}_n$ to represent each $u_i(k)$, which corresponds to the input of $\mathcal{D}$ in (\ref{eq:agree_functions_transfer}).
Then, the yielded vector $\vec{y}^{(i)}(k)={y}_0^{(i)}(k)\vec{e}_0+{y}_1^{(i)}(k)\vec{e}_1+\dots+{y}_n^{(i)}(k)\vec{e}_n$ in each local system $D^{(i)}$ is input to the agreement decision block $G_s^{(i)}$ in each node $i$ to derive the next decision value $s_i$.
As it is trivial to hold $s_i$ and not output it until round $k_f$ (for example $G_z^{(i)}(s_i(k),k)=s_i(k)\cdot \delta_{+k_f}[k]$ would do), the main problem is to solve $G_s^{(i)}$.

We note that in solving and realizing the local functions $G_s$ and $G_z$, no extra message is really needed to be exchanged in the system other than the ones being exchanged in the underlying broadcast system $\mathcal{D}$.
So the modularized system structure of $\mathcal{H}$ is free of the underlying network topologies.
In this way, all the broadcast solutions provided in this paper can be utilized as the underlying broadcast system $\mathcal{D}$ in $\mathcal{H}$.
The only added process is in the local functions $G_s$ and $G_z$, which requires very few resources.
And for the incomplete solutions, as the construction of the npc node set $P(T,\beta_0)$ for every $T$ is identical in the broadcast system and the agreement system, no extra effort is needed in constructing the agreement between the npc nodes with any $T$ during the execution of the system.

In the next subsection, we give a specific realization of the complete agreement system constructed upon a fully connected bipartite network.
The construction of agreement systems upon bipartite bounded-degree networks is straightforward.

\subsection{A specific solution for the BABi problem}
In \cite{RN4346}, a framework is presented for designing BA algorithms in general networks.
With the modularized structure shown in Fig.~\ref{fig:agreement_system2}, this framework can be applied to agreement systems based on the broadcast systems.
As we have extended the broadcast systems to bipartite networks, a sufficient solution for the BABi problem is straightforward.

To be compatible with \cite{RN4346} where the \emph{General} is regarded as a node in the network, we assume that each node $i\in V_A$ is initiated with an initial value $r_i$ by the \emph{General} (let suppose it as any a node in $V_B$) of the agreement system.
Then, all correct nodes (include $U_A$ and $U_B$) in the network are required to reach an agreement for this \emph{General} before a fixed number of rounds.
For this, in the most straightforward way, the top-rank \emph{Lieutenants} (the nodes in $V_A$) can act as the second-rank \emph{Generals} (the \emph{Generals} in $\mathcal{D}$) to initiate the independent broadcast primitives.
Alternatively, the top-rank \emph{Lieutenants} in $V_A$ can first distribute their initial values to the nodes in $V_B$.
Then, the nodes in $U_B$ would first do some fault-tolerant processing with the received initial values from $V_A$ in acquiring their initial values and then also act as the top-rank \emph{Lieutenants} (and act also as the second-rank \emph{Generals}) to initiate the independent broadcast primitives with their initial values.
The difference is that the problem of reaching agreement in $V_A$ can now be converted to the problem of reaching agreement in $V_B$ in the case of $|V_A|\gg |V_B|$.
The trick is that when the top-rank \emph{General} is correct, all correct nodes in $V_A$ would be consistently initialized, with which the nodes in $V_B$ would be consistently initialized too and thus the desired \emph{validity} property follows.
And when the top-rank \emph{General} is faulty, the following agreement running for $V_B$ can still maintain the desired \emph{agreement} property.
So here we take this alternative way.

For this, we assume $\vec r_A\equiv (s_0\cdot \vec 1_A+\vec{\omega}_0)\delta$ being the original input of $\mathcal{H}$, where $s_0\in \mathbb V$ and $\vec{\omega}_0\in \mathbb F^{n_A}$.
And the responding decision vector and noise vector are respectively $\vec{s}_B(k)=s_1(k)\vec{e}_1+\dots+s_{n_B}(k)\vec{e}_{n_B}$ and $\vec{\omega}^{(i)}_B(k)={\omega}_1^{(i)}(k)\vec{e}_1+\dots+{\omega}_{n_B}^{(i)}(k)\vec{e}_{n_B}$ for $k\in \mathbb N$.
The yielded vector of $\mathcal{D}$ in each node $i\in V$ is $\vec{y}^{(i)}=[\vec{y}_A^{(i)};\vec{y}_B^{(i)}]$ where $\vec{y}_A^{(i)}$ and $\vec{y}_B^{(i)}$ are respectively the yielded values in nodes of $V_A$ and $V_B$.
With our strategy, as no node in $V_A$ is allowed to initiate the broadcast primitive, the noisy signals $\hat{\vec{y}}_A^{(i)}(k)$ can be simply ignored in all correct nodes in $V$.
So only the noises ${\omega}_h^{(i)}$ for $h\in V_B$ need to be considered.
Without loss of generality, we can also set $\hat{\vec{s}}^{(i)}(-1)=\vec0$ for all $i\in V$.

Following \cite{RN4346}, $\mathcal{H}$ should maintain an invariant $\mathcal I$:
\begin{eqnarray}
\label{eq:agree_invariant}
\vec{s}(k)= \vec1 \lor \vec{s}(k)= \vec0 \lor f_k> k
\end{eqnarray}
where $f_k={\lVert \sum\nolimits_{t=0}^{k}\sum\nolimits_{i=1}^{n}|\vec{\omega}^{(i)}(t)|\rVert}$ is the number of actually faulty nodes up to the $k$th round.
As $f_k$ is bounded by $f$, $\vec{s}(k)= \vec1 \lor \vec{s}(k)= \vec0$ would hold when $k\geqslant f$.
To maintain this invariant, an intuition is that the noises generated by the static adversary can be eventually filtered out in the system $\mathcal{H}$, as these noises can be viewed as bounded noises in the $0$-norm space.
In our case, following the basic strategies in \cite{Toueg1987Fast}, for each node $j\in V_B$ we can set
\begin{eqnarray}
\label{eq:agree_gvb}
G_{s_B}(\vec{y}^{(j)}(k),k)=\left\{
\begin{aligned}
1 & & {{y}_0^{(j)}(k)=1 \land {\lVert \vec{y}_B^{(j)}(k) \rVert} > k}\\
0 & & \text{otherwise}
\end{aligned}
\right.
\end{eqnarray}
And for each node $i\in V_A$, we can simply set
\begin{eqnarray}
\label{eq:agree_gva}
G_{s_A}(\vec{y}^{(i)}(k),k)=\left\{
\begin{aligned}
1 & & {{\lVert \vec{y}_B^{(i)}(k) \rVert} \geqslant n_B-f_B}\\
0 & & {{\lVert \vec{y}_B^{(i)}(k) \rVert} \leqslant f_B}\\
v & & \text{otherwise}
\end{aligned}
\right.
\end{eqnarray}
where $v\in \mathbb V$ can be arbitrarily valued.

As $\mathcal{D}_{A,B}$ is a solution of the $(f_A,f_B)$-Byzantine broadcast problem, it satisfies the $1$-Heaviside and $1$-Dirac properties.
With the $1$-Heaviside property, for all correct nodes $i\in U_A$ and $j\in U_B$ we have
\begin{eqnarray}
\label{eq:agree_1h2}
s_j(k)=1 \to \forall t> k:{y}_j^{(i)}(t)=1 \\
\label{eq:agree_1h1}
{y}_j^{(i)}(k)=1 \to \exists 0 \leqslant t < k:\Delta s_j(t)=1
\end{eqnarray}
where $s_j(k)$ is the local decision value in node $j$ and ${y}_j^{(i)}(k)$ is the $\mathcal{D}_{A,B}$-yielded value of node $j$ in node $i$ at round $k$.
With the $1$-Dirac property, we have
\begin{eqnarray}
\label{eq:agree_2d}
\forall l,h,j\in V:{y}_j^{(l)}(k)\geqslant {y}_j^{(h)}(k-1)
\end{eqnarray}

Thus, following the proofs in \cite{Toueg1987Fast}, in round $k=f_B$, if a correct node $j\in V_B$ satisfies $s_j(k)=1$, with (\ref{eq:agree_functions_yield}) and (\ref{eq:agree_gvb}) it means ${\lVert \vec{y}_B^{(j)}(k) \rVert} > k=f_B$.
As $f_k\leqslant f_B$, there is at least one correct node $h\in U_B$ satisfies ${y}_h^{(j)}(k)=1$.
With (\ref{eq:agree_1h1}) and $s_h(-1)=0$, there exists $t\geqslant 0$ satisfies $t < k \land s_h(t)=1 \land s_h(t-1)=0$.
And with (\ref{eq:agree_gvb}) it means ${y}_0^{(h)}(t)=1 \land {\lVert \vec{y}_B^{(h)}(t) \rVert} > t$.
Now with (\ref{eq:agree_1h2}), $\forall l\in V_B:{y}_h^{(l)}(t+1)=1 \land {y}_0^{(l)}(t+1)=1$ holds.
Together with (\ref{eq:agree_2d}), it comes $\forall l\in V_B: {\lVert \vec{y}_B^{(l)}(t+1) \rVert} \geqslant {\lVert \vec{y}_B^{(h)}(t) \rVert}+1>t+1$.
Thus again with (\ref{eq:agree_functions_yield}) and (\ref{eq:agree_gvb}), $\forall l\in V_B:s_l(t+1)=1$ holds.
As $0 \leqslant t<k$, it arrives $\forall l\in V_B:s_l(k)=1$.
Thus, for all correct nodes $l,h\in V_B$, $s_l(f_B)=s_h(f_B)$ holds.
For a correct node $i\in U_A$, as all correct nodes in $V_B$ can be agreed at round $f_B$, with (\ref{eq:agree_functions_yield}) and (\ref{eq:agree_gva}), we have $\forall l\in V_A,h\in V_B:s_l(f_B+1)=s_h(f_B)$.

Thus, for reaching agreement, $\mathcal{H}$ can terminate at round $f_B+1$ (the first half of round $0$ is for converting the agreement to the smaller side of $K_{n_A,n_B}$ and the second half of round $f_B+1$ is for informing the nodes in $U_A$ with the final decision of the nodes in $U_B$).
As a correct node $i\in U$ with $s_i=1$ would be persistent with $s_i=1$ (or saying monotonically increasing), all nodes in $U$ can reach BA at round $f_B+1$.
Here this BABi solution is referred to as a \emph{Byzantine-lever} (or precisely as a \emph{BA-lever} here for reaching BA) of $G$, as the agreement in the \emph{heavier} side of the bipartite network $G$ can be converted to the agreement in the \emph{lighter} side of $G$.
In Fig.~\ref{fig:algo_agree}, we present the corresponding algorithm $\mathtt{BA\_LEVER}$ for reaching agreement on any \emph{General} in $V_B$ or equivalently the $n_A$ initial values in $V_A$ (where the \emph{General} in $V_B$ is just nominal).
This algorithm is built on the broadcast primitive with two basic functions.
The function $initiate(i,g,v)$ initiates the agreement for the nominal \emph{General} $g$ with the initial value $v$.
The function $agree(i,g,v)$ makes agreement decision on $(g,v)$ in node $i$.
Here we set $k_f=f_B+1$ for the \emph{General} in $V_B$.
As we take the (nominal) \emph{General} in $V_B$ and there can be at most $n_B$ parallel broadcast primitives during the execution of $\mathtt{BA\_LEVER}$, the required overall traffic is no more than $2n_A n_B^2 \lceil\log_2 {n_B}\rceil$ bits.

\alglanguage{pseudocode}
\algrenewcommand{\algorithmiccomment}[1]{\hskip1em//#1}
\begin{figure}[htbp]
\centering
\subfloat[For Node $i\in V_A$\label{algo:agree-A}]{
\begin{minipage}{.20\textwidth}
\centering
\begin{algorithmic}[1]
\Statex {\underline{$initiate(i,g,v)$}:}
    \State  $g_0:=g$;
    \State  $s_i:=0$;
    \State  $\vec{y}_B^{(i)}:=\vec 0$;
    \ForAll {$h\in V_B$}
    \State  $init(i,h,0)$;
    \State  $bcast(i,h,v,0)$;
    \EndFor
\Statex
\Statex
\Statex {\underline{on $accept(i,h,v,k)$}:}
    \State  $y_h^{(i)}:=1$;
    \If{${\lVert \vec{y}_B^{(i)} \rVert} > f_B$}
        \State  $s_i:=1$;
    \EndIf
\Statex
\Statex
\Statex {\underline{on end of round $k$}:}
    \If{$k=k_f$}
		\State  $agree(i,g_0,s_i)$;
    \EndIf
\end{algorithmic}
\end{minipage}}
\hfill
\subfloat[For Node $j\in V_B$\label{algo:agree-B}]{
\begin{minipage}{.27\textwidth}
\centering
\begin{algorithmic}[1]
\Statex {\underline{$initiate(j,g,0)$}:}
    \State  $g_0:=g$;
    ~  $s_j:=0$;
    \State  $\vec{y}_B^{(j)}:=\vec 0$;
    ~  $\vec{r}_A^{(j)}:=\vec 0$;
    \ForAll {$h\in V_B$}
    \State  $init(j,h,0)$;
    \EndFor
\Statex {\underline{on $recv(i,j,g_0,1,0)$}:}
    \State  $r_i^{(j)}:=1$;
    \If{${\lVert \vec{r}_A^{(j)} \rVert} > f_A$}
        \State  $bcast(j,j,1,0)$;
    \EndIf
\Statex {\underline{on $accept(j,h,v,k)$}:}
    \State  $y_h^{(j)}:=1$;
    \If{$s_j=0\land y_{g_0}^{(j)}=1\land{\lVert \vec{y}_B^{(j)} \rVert} > k$}
        \State  $s_j:=1$;
        ~  $bcast(j,j,1,k)$;
    \EndIf
\Statex {\underline{on end of round $k$}:}
    \If{$k=k_f$}
		\State  $agree(j,g_0,s_j)$;
    \EndIf
\end{algorithmic}
\end{minipage}}
\caption{The $\mathtt{BA\_LEVER}$ Algorithm.}\label{fig:algo_agree}
\end{figure}

\subsection{A general extension}
In general cases, the $G_s$ function can be extended to
\begin{eqnarray}
\label{eq:agree_gvgeneral}
G_{s}(\vec{y}^{(i)}(k),k)=\left\{
\begin{aligned}
1 & & {{y}_0^{(i)}(k)=1 \land k_0{\lVert \vec{y}^{(i)}(k') \rVert} \geqslant k'}\\
0 & & \text{otherwise}
\end{aligned}
\right.
\end{eqnarray}
with $k_0=\max\{k_H,k_\delta\}$ and $k'=k-k_0$.

\begin{theorem}
\label{theorem_general_extension}
If there is an $(\alpha,\mu)$-resilient $(k_H,k_\delta)$ broadcast system $\mathcal{D}$ upon $G$ with complexity $O(X)$ for a single broadcast, then $(\alpha,\mu)$-resilient agreement system $\mathcal{H}$ upon $G$ exists with complexity $O(fX)$ and termination time $k_f=k_0 \lceil \mu f  +1 \rceil $.
\end{theorem}
\begin{IEEEproof}
Firstly, with $\mathcal{D}$, the properties provided in (\ref{eq:agree_1h2}), (\ref{eq:agree_1h1}) and (\ref{eq:agree_2d}) now become
\begin{eqnarray}
\label{eq:agree_1h2_general}
s_j(k)=1 \to \forall t\geqslant k+k_H:{y}_j^{(i)}(t)=1 \\
\label{eq:agree_1h1_general}
{y}_j^{(i)}(k)=1 \to \exists 0 \leqslant t \leqslant k-k_H:\Delta s_j(t)=1 \\
\label{eq:agree_2d_general}
\forall l,h,j\in V:{y}_j^{(l)}(k)\geqslant {y}_j^{(h)}(k-k_\delta)
\end{eqnarray}
for $i,j\in P(T)$ with the specific $T$.
So, at round $k=k_f$, if there is $j\in P(T)$ satisfying $s_j(k)=1$, at least one node $h\in P(T)$ satisfies ${y}_h^{(j)}(k')=1$ with $k'=k-k_0$.
So with (\ref{eq:agree_1h1_general}), there exists $0\leqslant t \leqslant k'$ satisfies $s_h(t)=1 \land s_h(t-1)=0$.
And with (\ref{eq:agree_gvgeneral}) it means ${y}_0^{(h)}(t)=1 \land k_0{\lVert \vec{y}^{(h)}(t-k_0) \rVert} \geqslant t-k_0$.
Now with (\ref{eq:agree_1h2_general}), $\forall l\in P(T):{y}_h^{(l)}(t)=1 \land {y}_0^{(l)}(t+k_0)=1$ holds.
Together with (\ref{eq:agree_2d_general}), it comes $\forall l\in P(T): k_0{\lVert \vec{y}^{(l)}(t) \rVert} \geqslant k_0({\lVert \vec{y}^{(h)}(t-k_0) \rVert}+1)\geqslant t$.
Thus again with (\ref{eq:agree_functions_yield}) and (\ref{eq:agree_gvgeneral}), $\forall l\in P(T):s_l(t+k_0)=1$ holds.
As $0 \leqslant t\leqslant k'=k-k_0$, it arrives $\forall l\in P(T):s_l(k)=1$.
So $\forall i,j\in P(T): s_j(k)=s_i(k)$ holds.
For efficiency, at most $O(f)$ broadcast instances run in an agreement.
\end{IEEEproof}

\subsection{A discussion of efficiency}

As the multi-valued agreements can be reduced to the Boolean ones, here we only compare the efficiency of $\{0,1\}$ agreements.
And as is introduced earlier, here we mainly focus on the deterministic synchronous immediate BA solutions (complete and incomplete) without authenticated messages nor early-stopping optimization.

Firstly, for the $(\alpha,1)$-resilient agreements, we compare the $\mathtt{BA\_LEVER}$ algorithm with the classical optimal ones in Table~\ref{tab:complete}.
For time (round) efficiency, as $f=f_A+f_B$, the immediate agreement can be reached in bipartite networks with less than $f+1$ rounds, which can be better than the low-bound proved in \cite{FischerA} under freely allocated Byzantine nodes in fully connected networks.
And the added $1$ initial round is only for the BA-levers.
For communication complexity, the required overall traffic in executing the $\mathtt{BA\_LEVER}$ algorithm is bounded by $2n_A n_B^2\lceil\log_2 {n_B}\rceil\}$ bits, which is better than the results in \cite{Kowalski2013Synchronous} (at least $O(n^3\log n)$).
And when $n_A\gg n_B$ holds, required traffic can be nearly linear to $n_A$.
The number of messages in all point-to-point channels are also reduced to $n_A n_B$ during a round.
For computational complexity in a node, as there can be at most $n_B$ parallel broadcasts in $\mathtt{BA\_LEVER}$, each of which sums up at most $n_A$ Boolean values in a round, required computation and storage in a round are no more than $O(n_A n_B)$, which are also better than those of \cite{Kowalski2013Synchronous,RN4148} where the sizes and computations of dynamic trees are at least $\Omega (n^b)$ with $b>1$.
For network scalability, required node-degrees in nodes of $V_A$ and $V_B$ are respectively bounded by $n_B$ and $n_A$, which is better than $n-1$ in fully connected networks.
And required connections in the bipartite network are linear to $n_A$ when $n_A\gg n_B$, which is also better than that in fully connected networks.
For resilience, to tolerate $f_A$ and $f_B$ Byzantine nodes, it is required that $n=n_A+n_B\geqslant 3(f_A+f_B)+2=3f+2$, which only needs one more node in comparing with optimal resilience ($n\geqslant 3f+1$) in fully connected networks.

\begin{table}[htbp]
\caption{A comparison for the $(\alpha,1)$-resilient deterministic BA}
\label{tab:complete}       % Give a unique label
% For LaTeX tables use
\begin{tabular}{lll}
\hline\noalign{\smallskip}
efficiency &this paper                    &classical \\
\noalign{\smallskip}\hline\noalign{\smallskip}
time (rounds)           &$f_B+2$                            &$f+1$\\
traffic (overall bits)  &$O(n_A n_B^2 \log {n_B})$          &$O(n^3\log n)$\\
message (per round)     &$O(n_A n_B)$                       &$O(n^2)$\\
computation (per node)  &$O(n_A n_B)$                       &$O(n^2)$\\
storage (per node)      &$O(n_A n_B)$                       &$O(n^2)$\\
required node-degrees   &$(n_B,n_A)$                        &$n-1$\\
number of connections   &$n_B n_A$                          &$n(n-1)$\\
Byzantine resilience    &$(\frac{1}{3+\epsilon},\frac{1}{3+\epsilon})$  &$\frac{1}{3+\epsilon}$\\
\noalign{\smallskip}\hline
\end{tabular}
\end{table}

For the $(\alpha,\mu)$-resilient agreements with $\mu>1$, the required node-degree and connections remain the same as the ones in $\mathcal{D}$ under the same Byzantine resilience.
As the $G_s$ and $G_z$ functions require very few computations and storage resources, the efficiency mainly depends on the underlying broadcast system.
For time efficiency, the required rounds are $\max\{k_H,k_\delta\} \lceil \mu f +1 \rceil$ in \emph{worst cases}.
With \cite{Toueg1987Fast} it is easy to optimize the non-worst-cases with early-stopping \citep{Dolev1990Early}.
And for the \emph{worst cases}, if we can further refine the Heaviside and Dirac properties with $(k_H^{(min)},k_H^{(max)})$ and $(k_\delta^{(min)},k_\delta^{(max)})$ to some extent, the required rounds can be further reduced.

Note that, in comparing the efficiency of the BABi solutions with that of the classical ones, we assume that the failure rates and the deployed numbers of the nodes in two sides of the bipartite network can be significantly different.
This assumption largely comes from some real-world systems where the computing components (such as a huge number of sensors, actuators, embedded processors) and the communicating components (such as the customized Ethernet switches, directional antennas) are fundamentally different.
In the next section, we give several examples of such systems and show some special time efficiency gained in the bipartite solutions.

\section{Application with high assumption coverage}
\label{sec:app}
In applying the BABi solutions in practice, it is critical to show the gained networking, time, computation, and communication efficiencies are not at the expense of degraded system reliability.
This sometimes depends on the features of the concrete real-world systems.
Especially, the Byzantine resilience towards the two sides of the bipartite networks should be well configured according to concrete situations.
In this section, we discuss this problem with some possible applications.

\subsection{Applying to the WALDEN approach}
\subsubsection{Realization in WALDEN}
The WALDEN (Wire-Adapted Link-Decoupled EtherNet) approach is proposed in \cite{YuCOTS2021} for constructing distributed self-stabilizing synchronization systems and upper-layer synchronous applications upon common switched-Ethernet components.
Once the desired synchronization is reached (in a deterministically bounded time) in the WALDEN network, the solution provides a globally aligned user stage (called the TT-stage \citep{YuCOTS2021}) for performing upper-layer real-time synchronous activities such as TDMA communication and further executing round-based semi-synchronous protocols.
For example, in supporting fault-tolerant TDMA communication, following the basic fault-containment strategies in \cite{as6802}, the WALDEN switches can perform the corresponding temporal isolations for the incoming and outgoing traffic with pre-scheduled time-slots.

To apply the BABi solution in the WALDEN networks, here we assume a fully connected bipartite network $G_\mathtt{W}=(V_A\cup V_B,E)$ is composed of the advanced WALDEN end-systems $V_A$ and the WALDEN switches $V_B$.
For our basic purpose here, we assume the desired semi-synchronous communication rounds in the network $G_\mathtt{W}$ have been established in the TT-stages of the synchronized WALDEN system and the BABi protocols can be executed in a single TT-stage.
Concretely, we assume the TT-stage contains a sufficient number of communication rounds and each communication round contains one or several time-aligned slots for every correct node.
For simplicity, we assume each $\mathtt{BI\_BROADCAST}$ primitive runs in exclusive slots among the other ones.

With this, we show that the $\mathtt{BI\_BROADCAST}$ primitive simulated in $G_\mathtt{W}$ would not add significant computing burden nor processing delay on the WALDEN switches.
Firstly, as each WALDEN switch (WS) can perform temporal isolation (see \cite{YuCOTS2021} for details) for the fault-tolerant TDMA, it can also conveniently perform this for realizing the $D_{x_B}$ functions.
Concretely, in the globally aligned TT-stage, during each simulated synchronous round of the BABi protocol, the outgoing traffic in each WS can be isolated from the receivers before some specific condition being satisfied.
Namely, the \emph{out-gates} in the WA (wire-adapter) components of the WS would be closed by default in each slot during the simulation of the BABi protocol until a sufficient number of messages arrive in the same switch during the same slot.
For the corresponding $D_{x_B}$, the \emph{out-gates} of each WS would be opened in each slot if messages are detected in the \emph{in-watch} signals of at least $f_A+1$ distinct WA components in the same WS during the same slot.
Notice that the last transmitted message in these first arrival $f_A+1$ messages during a slot would pass the \emph{out-gates} of the WS without any extra delay, the communication delays between the ES nodes are not enlarged by the fault-tolerant processing in the WS nodes.

Obviously, for such strategies working, the first arrival messages in the $f_A+1$ distinct WA components should be processed sufficiently fast for at least one message of them being properly transmitted in all outgoing channels of the WS node.
We can see this is easily supported in the WALDEN scheme, as the physical-layer data-arriving signal of each message in each slot can be promptly delivered into the centralized DL (decoupled linker) component of the WS before the message entering the standard Ethernet switch or even being fully decoded in the WA components.
And the operations of the \emph{out-gates}, namely, open and close, can also be performed promptly with the minimal hardware modification proposed in \cite{YuCOTS2021}.
With this, as a WS node can definitely know which kind of messages is expected in each incoming channel during each slot with the statically scheduled TDMA rules, the WS node can even finish the core fault-tolerant processing (namely, the $D_{x_B}$ function) before the messages being decoded in the data-link-layer.
Similarly, the corresponding $D_{y_B}$ function can also be realized by accepting the simulated broadcast with the messages from at least $n_A-f_A$ distinct WA components.
And to fulfill all of these, the only added computation for the $\mathtt{BI\_BROADCAST}$ primitive is to count the number of WA components with nonzero data-arriving signals in each slot, which is trivial in considering just the decoding of messages in the data-link-layer of Ethernet.

\subsubsection{Efficiency and assumption coverage}
In running the overall $\mathtt{BA\_LEVER}$ algorithm in the WALDEN network, when the agreement system is initialized, no extra message needs to be sent during each communication round other than the ones being sent in simulating the $\mathtt{BI\_BROADCAST}$ primitives.
Thus, the required computation and communication in each slot remain unchanged, which is optimal in the deterministic settings.

For the overall efficiency, one arguable aspect might be the exclusive usages of the slots in each $\mathtt{BI\_BROADCAST}$ primitive.
Namely, as there could be $n_B$ temporal-exclusive $\mathtt{BI\_BROADCAST}$ primitives run in the overall algorithm, the duration of each communication round is not independent of $n$.
Notice that, however, the traffic required in each channel during each communication round is just $O(n)$ bits in $G_\mathtt{W}$, which is better than the optimal $O(n\log n)$ bits if all the messages in a communication round are allowed to be sent simultaneously and silently \citep{Silence2018}.
The main reason is that the TDMA communication can eliminate the expense on the identifiers of the \emph{Generals} in the expense of strictly separated time-slots.
Thus, the overall time-efficiency of the $\mathtt{BA\_LEVER}$ algorithm in $G_\mathtt{W}$ actually depends on the precision of the underlying synchronization scheme.
To this, as the spacing between adjacent time-slot in $G_\mathtt{W}$ is lower-bounded by $O(\log n)$ (being investigated in \citep{Gradient2004}), the overall communication time can be at the same order of the communication protocols with the \emph{silence} strategies \citep{Silence2018}.

For system assumption coverage, the componential failure rates of the ES nodes and the WS nodes might be both high.
Nevertheless, by assuming the independence of the componential failures of different nodes in the WALDEN networks \citep{BPLSNHSAC}, it is very unlikely that there would be more than $n_A/3$ faulty ES nodes or $n_B/3$ faulty WS nodes during the execution of the system when $n_A$ and $n_B$ are sufficiently large.
For example, with allowing that the componential failure rates per hour of both the ES nodes and the WS nodes are $p=10^{-3}$, when $f_A$ and $f_B$ are larger than $1$, a lower bound of the system assumption coverage $R_\mathtt{W}$ (per hour, the same below) can be represented as \citep{BPLSNHSAC}
\begin{eqnarray}
\label{eq:sac_WALDEN}
R_\mathtt{W}\geqslant 1-\sum_{i=f_A+1}^{n_A}{\tbinom{n_A}{i} p^{i}(1-p)^{n_A-i}}-\nonumber\\
\sum_{i=f_B+1}^{n_B}{\tbinom{n_B}{i} p^{i}(1-p)^{n_B-i}}\nonumber\\
\approx1-\sum_{i=f_A+1}^{n_A}{\sqrt{\frac{1}{2\pi i}}(\frac{epn_A}{i})^i (1-p)^{n_A-i}}-\nonumber\\
\sum_{i=f_B+1}^{n_B}{\sqrt{\frac{1}{2\pi i}}(\frac{epn_B}{i})^i (1-p)^{n_B-i}}\nonumber\\
\end{eqnarray}
So when $epn_A<1$ and $epn_B<1$ hold, $R_\mathtt{W}$ can be estimated as
\begin{eqnarray}
\label{eq:sac_WALDEN2}
R_\mathtt{W}\geqslant1-{\sqrt{\frac{1}{2\pi (f_A+1)}}(\frac{epn_A}{f_A+1})^{f_A+1}}/{(1-epn_A)}-\nonumber\\
{\sqrt{\frac{1}{2\pi (f_B+1)}}(\frac{epn_B}{f_B+1})^{f_B+1}}/{(1-epn_B)}
\end{eqnarray}

For example, when we take $n_A=n_B=10$ and $f_A=f_B=3$, the system assumption coverage $R_\mathtt{W}$ would be larger than $1-2\frac{(2.7\times 10^{-2})^4}{4^5(1-2.7\times 10^{-2})}>1-10^{-9}$.
So, as the provided $\mathtt{BA\_LEVER}$ algorithm can tolerate $f_A$ and $f_B$ Byzantine nodes in $V_A$ and $V_B$ respectively, the system failure rate would be lower than $10^{-9}$ per hour in this case, providing that the componential failures of different nodes are independent.

Lastly, in these bipartite networks, if the componential failure rates in one side are significantly larger than those of the other side, the system assumption coverage can be maintained by deploying more nodes in the former side.
In doing this, the execution time can be remained as low as possible by applying the Byzantine-levers.
\subsection{Some other possible applications}
Although the solutions provided for the arbitrarily connected networks can be with the widest application, the ones provided for the bipartite networks can be with special usefulness, as the two sides of the bipartite networks can often be interpreted as some cooperatively-coupled operations.

For example, although wireless communication technologies have apparent advantages in omnidirectional connections, they are often vulnerable to electromagnetic attacks.
The restricted communication connections upon bipartite topologies, on the contrary, do not require any communication between the specific objects deployed in the same layer.
This adds extra resilience to some emerging ad-hoc systems, such as large-scale unmanned aerial vehicle (UAV) formations, in the \emph{open} or even \emph{antagonistic} environment where the omnidirectional wireless communication might be a disadvantage.
For example, it is reported that large-scale UAV accidents are often caused by some large-range communication failures that are unlikely independent with each other.
In this context, directional communication schemes (with redundant ground stations or just several flocks of UAVs) can be deigned to make the physical communication as secure as possible.
With this, even if the communication system is under attack at the specific directions, the attackers can be easily detected and fast located with low-cost radars and other directional guardians without omnidirectional scanning.
Meanwhile, in considering the system assumption coverage without external attacks, although the UAV units might be with higher componential failure rates than the ground stations, they can often be deployed with a great number.
With this, it is very unlikely that more than a third of the UAV units in a flock fail in the independent way \citep{BPLSNHSAC}.
Also, with applying the Byzantine-levers, the execution time of the BA routine can be reduced with the well-guarded redundant ground stations.

For another example, consider the near-earth space communication networks where the LEO satellites (on low earth orbits) communicate with the ground stations (on the earth), MEO satellites (on higher orbits), or even the stationary GEO satellites (on the geostationary orbits \citep{CLARKE19663}).
As is limited by the power supply, communication bandwidth, antenna directivity, and so on, the communication channels provided between the nodes of the network are often severely restricted.
In this situation, bipartite and even multi-layer space communication networks can be further explored with allowing different componential failure rates in different layers.
It is easy to see that the provided broadcast and agreement solutions upon bipartite networks can be extended to such networks with similar resilience.
For efficiency, one of the most significant problems in space communication networks is the enlarged propagation delays.
With the fault-tolerant solutions provided for bipartite networks here, the two-way ground-space communication round can be utilized by the nodes of each layer to simulate a multi-phase fault-tolerant operation.
Meanwhile, the layer with the highest reliability can provide the fastest BA for all the other layers with the Byzantine-levers.

In the same vein, the bipartite solutions might also find their application in some future IoT systems where the communication infrastructures are often heterogeneous, i.e., the communicating components provided in the ever-changing real-world technologies (the legacy fast Ethernet, Gbit Ethernet, WiFi, Optical fiber networks, etc) are sometimes not able to communicate directly but can be connected in two hops with common computing components that equipped with multiple embedded IoT communicating modules.
In this context, as the communicating components can often be deployed far away from the edge computing nodes, the failure rates of the computing components and the communicating components should be separately considered, too.

In a broader meaning, the application of the BABi solutions discussed so far are all about the coupling of computation and communication.
It is interesting to find that BFT computation and BFT communication naturally show some duality in the context of BA.
Namely, we can say that BFT computation is recursively relies on BFT communication while BFT communication is recursively relies on BFT computation.
In this vein, the BA problem can be more naturally represented with bipartite graphs rather than complete graphs.
Generally, any cooperatively-coupled operations with finite sizes can be represented and simulated upon bipartite graphs.
In considering a fraction of such operations being under the full control of an adversary, the BABi solutions can be viewed as general winning strategies in playing the game with cooperatively-coupled operations.

\section{Conclusion}
\label{sec:Conclusion}
In this paper, we have investigated the BA problem upon bipartite networks and have provided several solutions by mainly extending the relay-based broadcast system \citep{SrikanthSimulating} and integrating the general broadcast system with a general framework for agreement \citep{RN4346}.

Firstly, some necessary conditions for satisfying the properties of the broadcast systems are formally identified.
Then, a general strategy of constructing the relay-based broadcast systems is given.
With this, the original relay-based broadcast system \citep{SrikanthSimulating} is extended to fully connected bipartite networks.
Then, the broadcast systems upon bipartite bounded-degree networks are also provided and discussed.
Meanwhile, a simulation of the a.e. BFT solutions upon butterfly networks is provided in the corresponding bipartite networks.
A basic property for developing BFT solutions upon bipartite expanders is also given by extending the classical result provided only for non-bipartite expanders.
Then, with the modularized system structure $\mathcal{H}$, the developed broadcast systems are employed as the building blocks in constructing efficient agreement systems.

To show the special efficiency gained in the bipartite networks, algorithms of the broadcast system and agreement system upon complete bipartite networks are provided.
The $\mathtt{BI\_BROADCAST}$ algorithm requires no more than $2n_A n_B \lceil\log_2 n\rceil$ bits for reaching Byzantine broadcast with $n$ distinct \emph{Generals}.
And the algorithm $\mathtt{BA\_LEVER}$ requires no more than $f_B+2$ rounds for reaching immediate BA, during which no more than $2n_A n_B^2 \lceil\log_2 {n_B}\rceil$ bits is required.
We have shown that with programmable communicating components, the complexity of messages, storages, computation, time, and the requirements on the number of connections and node-degrees of the BABi algorithm can be all better than the classical optimal BA algorithms upon fully connected networks if only the up to $f$ Byzantine nodes could be divided into $f_A$ and $f_B$ faulty end systems and switches with high assumption coverage.
In the given examples, we have shown that this assumption is reasonable.
Meanwhile, the BA-levers can provide leverage for converting the agreement of one side of the network to the other side.
This is done by exploiting a seldom noticed fact (or saying the given \emph{fulcrum}) that the failure rates of the computing components and the communicating components may be significantly different.
As more and more real-world communication networks are comprised of programmable communicating components (such as the widely deployed embedded systems, FPGA systems, software-defined networking products, wireless stations, directional antennas, software-defined radio, or even communication satellites), the failure rates of the nodes in different layers of the networks should better be considered separately.
In this trend, the Byzantine-levers can manifest themselves in prominent leverage effects.

Despite the merits, the results shown in this paper are rather heuristic than being ready for ultimate real-world applications.
From the practical perspective, these results can be further optimized in many ways.
Firstly, for the complete solutions, practical communication networks like the fat-trees \citep{Leighton1992On} may have multi-layer architectures (can still be represented under the bipartite topologies).
So the solutions upon fully connected bipartite network can be extended to multi-layer (bipartite) networks with well-distributed resilience.
Secondly, in the incomplete solutions, multi-scale BFT systems \citep{BPLSNHSAC} can be better explored with the provided basic properties of bipartite expanders.

\bibliographystyle{IEEEtran}
\bibliography{IEEEabrv,BBA}

\end{document}